\documentclass{article}

\usepackage{microtype}
\usepackage{graphicx}
\usepackage{booktabs} 
\usepackage{subfig}
\usepackage{algorithmic}
\usepackage{algorithm}
\usepackage{amstext}
\usepackage{enumitem}
\usepackage[T1]{fontenc}
\usepackage{balance}
\usepackage{amsmath}
\usepackage{tikz}
\newcommand*\circled[1]{\tikz[baseline=(char.base)]{
            \node[shape=circle,draw,inner sep=.6pt] (char) {#1};}}

\usepackage{hyperref}

\usepackage[accepted]{mlsys2020}

\mlsystitlerunning{Distributed Hierarchical GPU Parameter Server for Massive Scale Deep Learning Ads Systems}

\begin{document}

\twocolumn[
\mlsystitle{\vspace{-0.1in}Distributed Hierarchical GPU Parameter Server for\\ Massive Scale Deep Learning Ads Systems\vspace{-0.1in}}

\begin{mlsysauthorlist}
\mlsysauthor{Weijie Zhao$^1$, Deping Xie$^2$, Ronglai Jia$^2$, Yulei Qian$^2$, Ruiquan Ding$^3$, Mingming Sun$^1$, Ping Li$^1$}{}
\\

$ ^1$ Cognitive Computing Lab, Baidu Research \\
$ ^2$ Baidu Search Ads (Phoenix Nest), Baidu Inc. \\
$ ^3$ Sys. \& Basic Infra., Baidu Inc. \\
{\{weijiezhao, xiedeping01, jiaronglai, qianyulei, dingruiquan, sunmingming01, liping11\}@baidu.com}\\
\end{mlsysauthorlist}


\vskip 0.3in

\begin{abstract}
Neural networks of ads systems usually take input from multiple resources, e.g., query-ad relevance, ad features and user portraits. 
These inputs are encoded into one-hot or multi-hot binary features, with typically only a tiny fraction of nonzero feature values per example. 
Deep learning models in online advertising industries can have terabyte-scale parameters that do not fit in the GPU memory nor the CPU main memory on a computing node. 
For example, a sponsored online advertising system can contain more than $10^{11}$ sparse features, making the neural network a massive model with around 10 TB parameters. 
In this paper, we introduce a distributed GPU hierarchical parameter server for massive scale deep learning ads systems. We propose a hierarchical workflow that utilizes GPU High-Bandwidth Memory, CPU main memory and SSD as 3-layer hierarchical storage. All the neural network training computations are contained in GPUs. 
Extensive experiments on real-world data confirm the effectiveness and the scalability of the proposed system. A 4-node hierarchical GPU parameter server can train a model more than 2X faster than a 150-node in-memory distributed parameter server in an MPI cluster. In addition, the price-performance ratio of our proposed system is 4-9 times better than an MPI-cluster solution.
\vspace{-0.08in}
\end{abstract}

]

\printAffiliationsAndNotice{} 

\section{Introduction}\label{sec:introduction}

\begin{figure}[h!]
\centering
	\includegraphics[width=.46\textwidth]{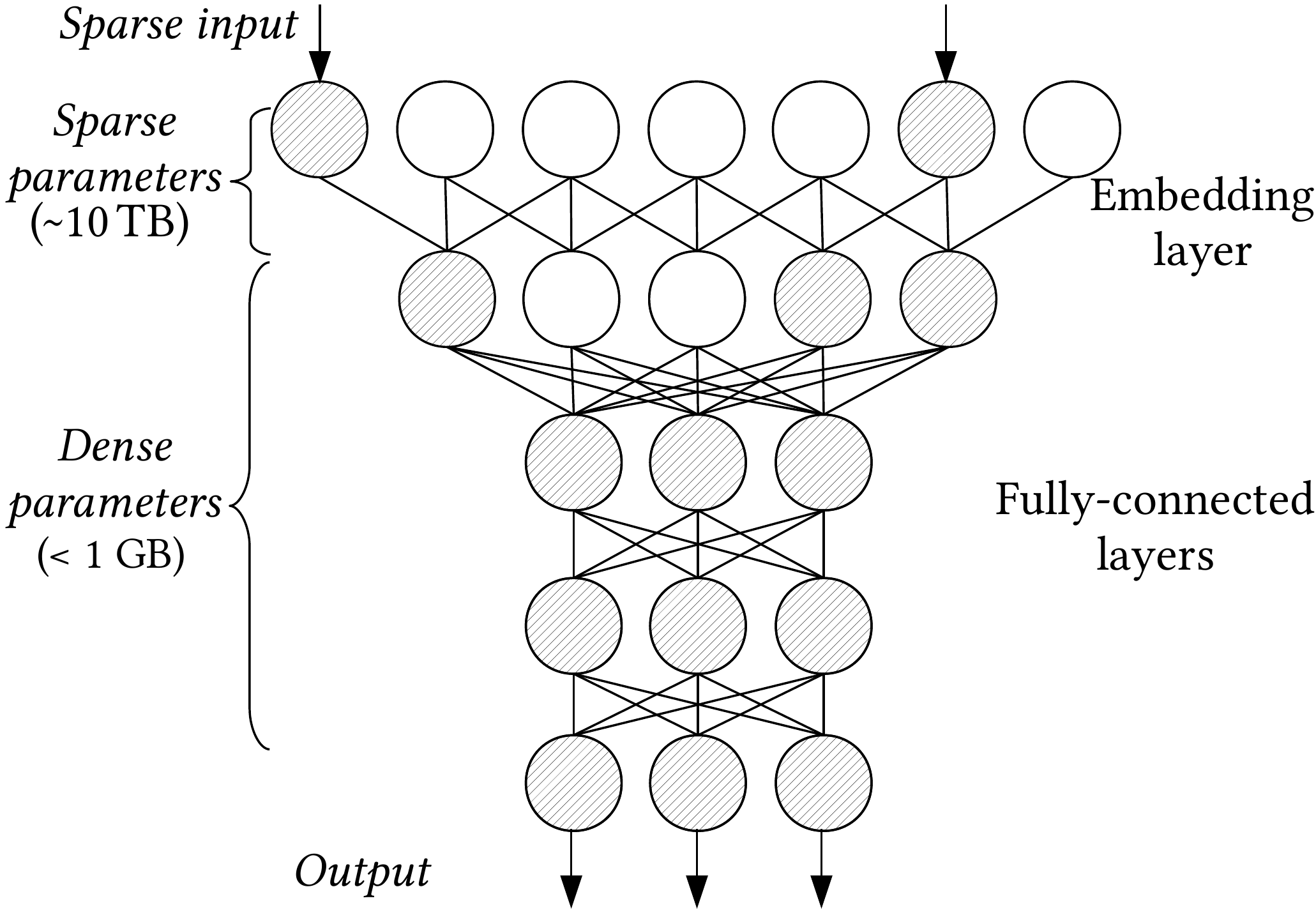}
    \vspace{-0.1in}
  	\caption{A visual illustration of CTR prediction network. The input data are sparse---only a tiny proportion of features are non-zero. The parameters of hatched neurons are referenced by~the~input.}
  	\label{fig:ctr-network}
    \vspace{-0.25in}
\end{figure}

Baidu Search Ads (a.k.a. ``Phoenix Nest'') has been successfully using ultra-high dimensional input data and ultra-large-scale deep neural networks for training CTR (Click-Through Rate) models since 2013~\cite{Proc:Fan_KDD19}.  Sponsored online advertising produces many billions of dollar revenues for online ad publishers such as Baidu, Bing, and Google. The task of CTR prediction~\cite{Article:Broder_2002,
fain2006sponsored,Proc:Fan_KDD19,Proc:Zhao_CIKM19} plays a key role to determine the best ad spaces allocation because it directly influences user experience and ads profitability. CTR prediction takes input from multiple resources--e.g., query-ad relevance, ad features, and user portraits--then estimates the probability that a user clicks on a given ad. 
These inputs are typically one-hot/multi-hot binary features with only a tiny fraction of nonzero feature values per example. The high-dimensional sparse input data first go through an embedding layer to get a low-dimensional embedding; then the embedded results are connected by fully-connected layers. A real-world sponsored online advertising system can contain more than $10^{11}$ sparse features, making the neural network a massive model with around 10 TB parameters. Figure~\ref{fig:ctr-network} provides an illustration of CTR prediction network. 

\subsection{CTR Prediction Based on MPI-Cluster Solution}

Since 2013, the standard practice in Baidu Search Ads has been using an MPI-based solution for training massive-scale deep learning CTR prediction.  Before that, Baidu had adopted a distributed logistic regression (LR) CTR model and distributed parameter server around 2010.  The MPI solution partitions the model parameters across computing nodes (e.g., 150 nodes). Each node obtains a batch of training data streamed from a distributed file system. The node pulls the required parameters from other nodes and computes the gradients. The gradients are pushed to the nodes that maintain the corresponding parameters through MPI communications. Baidu's MPI solution and distributed parameter server was also summarized in~\cite{li2014scaling} (while the first author in that paper was with Baidu Inc.).

We should mention that this paper focuses on the CTR models in the final stage of Baidu's ads system. In the earlier stage of the pipeline, Baidu has been extensively using recent advancements on approximate near neighbor search (ANNS) and maximum inner product search (MIPS)~\cite{Proc:Fan_KDD19,Proc:Zhou_NeurIPS19,Proc:Zhao_ICDE20,Proc:Tan_WSDM20} to improve the quality of ads recalls, which is a separate important task in the full pipeline of the ads system. 

\subsection{Drawbacks of MPI-Cluster Solution}
CTR model researchers and practitioners train the massive-scale model rapidly to test and verify their new ideas. However, it is impractical to reserve a large-scale computing cluster for everyone---the hardware maintenance and the energy cost of a large-scale computing cluster is pricey. Besides, further scaling up the training speed is challenging---in this large-scale distributed setting, communication and synchronization cost between computing nodes limits the training performance.
We seek to explore other directions to optimize the training system for massive-scale ads models.

\subsection{GPU Parameter Server for CTR Prediction}


Recently, GPUs attract broad attention as a cost-efficient machine learning accelerator. However, it is not straightforward to directly employ GPUs on the CTR prediction problem due to the limited GPU memory and excessive CPU-GPU data communications. 
This amount of parameters exceeds the amount of overall volume of GPU High-Bandwidth Memory (HBM) of any single computing node, and any practical computing cluster.
GPUs are considered inefficient to train the model that does not fit in the GPU memory~\cite{chilimbi2014project}. Prior efforts~\cite{Geeps,Proc:Zhao_CIKM19} cache the parameters in the CPU main memory and still require time-consuming data transferring between CPU and GPU to update parameters.
We notice that in the massive-scale ads system, although the entire model is huge, the number of non-zero features per example is small. 
For example, in Figure~\ref{fig:ctr-network}, the referenced parameters for the given input are hatched. The sparse input has $2$ non-zeros: the $1^{\textit{st}}$ and the $6^{\textit{th}}$ features. For the parameters in the embedding layer, only a subset of them is used and will be updated for this sparse input (we call them ``sparse parameters'' for convenience). Following that, the dense parameters in the fully-connected layers are all referenced. The CTR prediction neural network usually has at most a few million dense parameters---the memory footprint of dense parameters is small (around a few hundred megabytes).
The working parameters--the parameters that are referenced in the current batch input--can fit in the GPU memory. It allows us to distribute the working parameters in the GPU memory and accelerate the training with GPUs.

In this paper, we propose a hierarchical parameter server that builds a distributed hash table across multiple GPUs and performs direct inter-GPU communications to eliminate the CPU-GPU data transferring overhead. Moreover, the hierarchical design on top of SSDs with an in-memory cache also enables us to train a large-scale out-of-main-memory deep neural network on distributed GPU environments.

\textbf{Challenges \& approaches.}
There are three major challenges in the design of the hierarchical parameter server. 1) The first challenge is to construct an efficient distributed GPU hash table to store the working parameters in multiple GPUs in multiple computing nodes. We design intra-node GPU communications and collective communication strategies among multiple nodes to synchronize the parameters across all GPUs.
Remote direct memory access protocols are employed to enable direct peer-to-peer GPU communications without involving CPUs.
2) The second challenge is to devise a data transferring and caching mechanism to keep working parameters in memory. Data transferring and SSD I/Os bandwidth are relatively slow compared with the deep neural network training on GPUs. Without a carefully designed architecture, excessive network and disk I/Os can be a dominant performance slowdown.
We propose a 4-stage pipeline that overlaps inter-node network communications, SSD I/Os, CPU/GPU data transferring and GPU computations. All stages are optimized to match the inevitable GPU computations---the pipeline hides the I/O latency.
3) The third challenge is to effectively organize the materialized parameters on SSDs. The I/O granularity of SSDs is a block, while the training parameters we load is in key-value granularity. This mismatch incurs I/O amplification~\cite{lu2017wisckey}---unnecessary data are read from SSDs to load required parameters. In contrast, reading and writing contiguous addresses on SSDs are faster than random accesses---larger block size yields better disk I/O bandwidth.
We present a file-level parameter management strategy to write updates in batches as new files. A file compaction thread runs in the background to regularly merge files that contain a large proportion of stale parameters into new files.

\textbf{Contributions.}
The technical contributions we make in this paper can be summarized as follows:
\begin{itemize}[topsep=1pt,itemsep=1pt,partopsep=1pt, parsep=1pt, leftmargin=*]
\item We present a hashing method (OP+OSRP) to reduce CTR models. We observe that combining OP+OSRP with DNN-based models can replace the original LR-based models. However, for DNN-based CTR models, the hashing method will hurt the prediction accuracy to the extent that the revenue would be noticeably affected (Section~\ref{sec:hash}).
\item We introduce the architecture of a distributed hierarchical GPU parameter server (Section~\ref{sec:overview}). 
The hierarchical design employs the main memory and SSDs to store the out-of-GPU-memory and out-of-main-memory parameters when training massive-scale neural networks.
As far as we know, it is the first distributed GPU parameter server for the terabyte-scale sparse input deep learning.
\item We propose a novel HBM parameter server that keeps the working parameters in distributed GPU HBMs. The HBM parameter server enables GPU worker threads to efficiently fetch/update neural network parameters and to directly synchronize parameters across multiple GPU cards from multiple computing nodes (Section~\ref{sec:hbm-ps}).
\item We show a CPU main memory parameter server that prefetches and buffers the parameters of the future training batches. The memory parameter server hides the data loading and parameter transfer latency (Section~\ref{sec:mem-ps}).
\item We present an SSD parameter server that materializes the out-of-main-memory parameters in files. Parameter updates are written in batches onto SSDs as new files to fully utilize the I/O bandwidth. Regular compaction operations are performed to bound the disk usage (Section~\ref{sec:ssd-ps}).
\item We perform an extensive set of experiments on $5$ real CTR prediction models/datasets and compare the results with an MPI cluster training framework in the production environment (Section~\ref{sec:experiments}).
The 4-node distributed hierarchical GPU parameter server is 1.8-4.8X faster than the MPI solution.
The price-performance ratio of our proposed system is 4-9 times better than the MPI solution.
\end{itemize}

\vspace{-0.1in}
\section{Prior Effort: Hashing Methods for Reducing CTR Models}\label{sec:hash}

It is natural to ask why one could not simply use a good hashing method to reduce the model size.  In this section,  we report a brief summary of our prior efforts back in 2015 for developing hashing algorithms to reduce the size of CTR models, with the hope that hashing would not hurt accuracy. 

Hashing algorithms are popular these days~\cite{Proc:Weinberger_ICML09,Proc:HashLearning_NIPS11}. We had spent serious efforts to develop effective hashing methods suitable for  CTR models. The results are both encouraging and discouraging. Here we would like to  make a comment  that there is a difference between academia research and industrial practice.  In academia research, we often take it as a good result if, for example, we are able to achieve a 10-fold model reduction by losing only $1\%$ of accuracy. For commercial search, however, we often cannot  afford to lose an accuracy of $0.1\%$ since that would result in a noticeable decrease~in~revenue.

\textbf{One permutation + one sign random projection.}
In the course of this research, we had tried many versions of hashing. We eventually recommend the method called ``one permutation + one sign random projection (OP+OSRP)''. As the name suggests, our idea was inspired by several   well-known hashing methods, e.g.,~\cite{Article:Goemans_JACM95,Proc:Charikar_STOC02,Article:Charikar_2004,Article:Cormode_05,Proc:Weinberger_ICML09,Proc:HashLearning_NIPS11,Proc:Li_Owen_Zhang_NIPS12,Proc:Shrivastava_UAI14,Proc:Li_NeurIPS19_BCWS}.

We assume the dimensionality of the binary training data  is $p$. OP+OSRP consists of the following key steps:

\begin{enumerate}[topsep=1pt,itemsep=1pt,partopsep=1pt, parsep=1pt, leftmargin=*]
\item  Permute the $p$ columns (only once). This can be very efficiently done via standard 2U or 4U hashing.
\item  Break the permuted $p$ columns uniformly into {$k$} bins.
\item  Apply independent random projection in each bin, for example, in the first bin, we let $z = \sum_{i=1}^{p/k} x_i r_i$, where we sample $r_i \in \{-1, \ +1\} \text{ w. p. } 1/2$.
\item  Store the sign of $z$ and expand it into a vector of length 2: {[0 \  1]} \text{ if } z >0, {[1 \  0]} \text{ if } z < 0 , {[0 \  0]} \text{ if } z = 0. The hashed data will be binary  in {$2k$} dimensions. This step is  different from prior research. This way, we still obtain binary features so that we do not need to modify the  training algorithm (which is coded efficiently for~binary~data).
\end{enumerate}

OP+OSRP is very efficient (essentially by touching each nonzero entry once) and can naturally handle sparse data (i.e., many bins will be empty since $k$ cannot be small). 

\textbf{Experimental results.}
This set of experiments was conducted in 2015 using 3 months of sponsored ads click data from web search and 3 months of data from image search.  As one would expect, web search brings in majority of the revenue. Image search is fun and important but its revenue is only a tiny fraction of the revenue from web search. Understandably, lots of computing/storage resources have been allocated for building CTR models for web search.

\begin{table}[htbp]
\caption{OP+OSRP for Image Search Sponsored Ads Data}
\begin{center}{
\small
{\begin{tabular}{l r c c c c c }
\hline \hline
      & \# Nonzero Weights & Test AUC \\
\hline
Baseline LR &31,949,213,205 &0.7112 \\
Baseline DNN &  &0.7470\\
Hash+DNN (k = $2^{34}$) & 6,439,972,994 &0.7407\\
Hash+DNN (k = $2^{23}$) & 3,903,844,565&0.7388\\
Hash+DNN (k = $2^{22}$) & 2,275,442,496&0.7370\\
Hash+DNN (k = $2^{31}$) & 1,284,025,453&0.7339\\
Hash+DNN (k = $2^{30}$) & 707,983,366&0.7310\\
Hash+DNN (k = $2^{29}$) & 383,499,175&0.7278\\
Hash+DNN (k = $2^{28}$) & 203,864,439&0.7245\\
Hash+DNN (k = $2^{27}$) & 106,824,123&0.7208\\
Hash+DNN (k = $2^{26}$) & 55,363,771&0.7175\\
Hash+DNN (k = $2^{25}$) & 28,479,330&0.7132\\
Hash+DNN (k = $2^{24}$) & 14,617,324&0.7113
\\\hline\hline
\end{tabular}}
}
\end{center}\label{tab_ImageSearch}
\end{table}

Table~\ref{tab_ImageSearch} summarizes the experimental result for applying OP+OSRP on image search ads data. There are several important observations from the results:
\begin{itemize}[topsep=1pt,itemsep=1pt,partopsep=1pt, parsep=1pt, leftmargin=*]
\item Compared to the baseline LR model, the DNN model very substantially improves the AUC. This is basically the justification of adopting DNN models for CTR prediction. 
\item Hashing reduces the accuracy. Even with $k=2^{34}$, the test AUC is dropped by $0.7\%$. 
\item Hash+DNN is a good combination for replacing LR. Compared to the original baseline LR model, we can reduce the number of nonzero weights from 31B to merely 14.6M without affecting the accuracy. 
\end{itemize}

Table~\ref{tab_FengCao} summarizes the experiments on web search ads data. The trend is essentially similar to Table~\ref{tab_ImageSearch}. The main difference is that we cannot really propose to use Hash+DNN for web search ads CTR models, because that would reduce the accuracy of current DNN-based models and consequently would affect the revenue for the company.

\begin{table}[ht]
\vspace{-.4cm}
\caption{OP+OSRP for Web Search Sponsored Ads Data}
\begin{center}{
\small
{\begin{tabular}{l r c c c c c }
\hline \hline
       &\# Nonzero Weights  & Test AUC\\
\hline
Baseline LR &199,359,034,971 &0.7458 \\
Baseline DNN &  &0.7670\\
Hash+DNN (k = $2^{32}$) & 3,005,012,154 &0.7556\\
Hash+DNN (k = $2^{31}$) & 1,599,247,184 &0.7547\\
Hash+DNN (k = $2^{30}$) & 838,120,432 &0.7538\\
Hash+DNN (k = $2^{29}$) & 433,267,303 &0.7528\\
Hash+DNN (k = $2^{28}$) & 222,780,993 &0.7515\\
Hash+DNN (k = $2^{27}$) & 114,222,607 &0.7501\\
Hash+DNN (k = $2^{26}$) & 58,517,936 &0.7487\\
Hash+DNN (k = $2^{24}$) & 15,410,799 &0.7453\\
Hash+DNN (k = $2^{22}$) & 4,125,016&0.7408
\\\hline\hline
\end{tabular}}
}
\end{center}\label{tab_FengCao}
\vspace{-0.1in}
\end{table}

\textbf{Summary.}
This section summarizes our effort on developing effective hashing methods for ads CTR models. The work was done in 2015 and we had never attempted to publish the paper. The proposed algorithm, OP+OSRP, actually still has some novelty to date, although it obviously combines several previously known ideas. The experiments are exciting in a way because it shows that one can use a single machine to store the DNN model and can still achieve a noticeable increase in AUC compared to the original (large) LR model. However, for the main ads CTR model used in web search which brings in the majority of the revenue, we observe that the test accuracy is always dropped as soon as we try to hash the input data. This is not acceptable in the current business model because even a $0.1\%$ decrease in AUC would result in a noticeable decrease in revenue.

Therefore, this report helps explain  why we introduce the distributed hierarchical GPU parameter server in this paper to train the massive scale CTR models, in a lossless fashion.

\vspace{-0.05in}

\section{Distributed Hierarchical Parameter Server Overview}\label{sec:overview}

\begin{figure}[htbp]
	\hspace*{-.15in}
	\includegraphics[width=.51\textwidth]{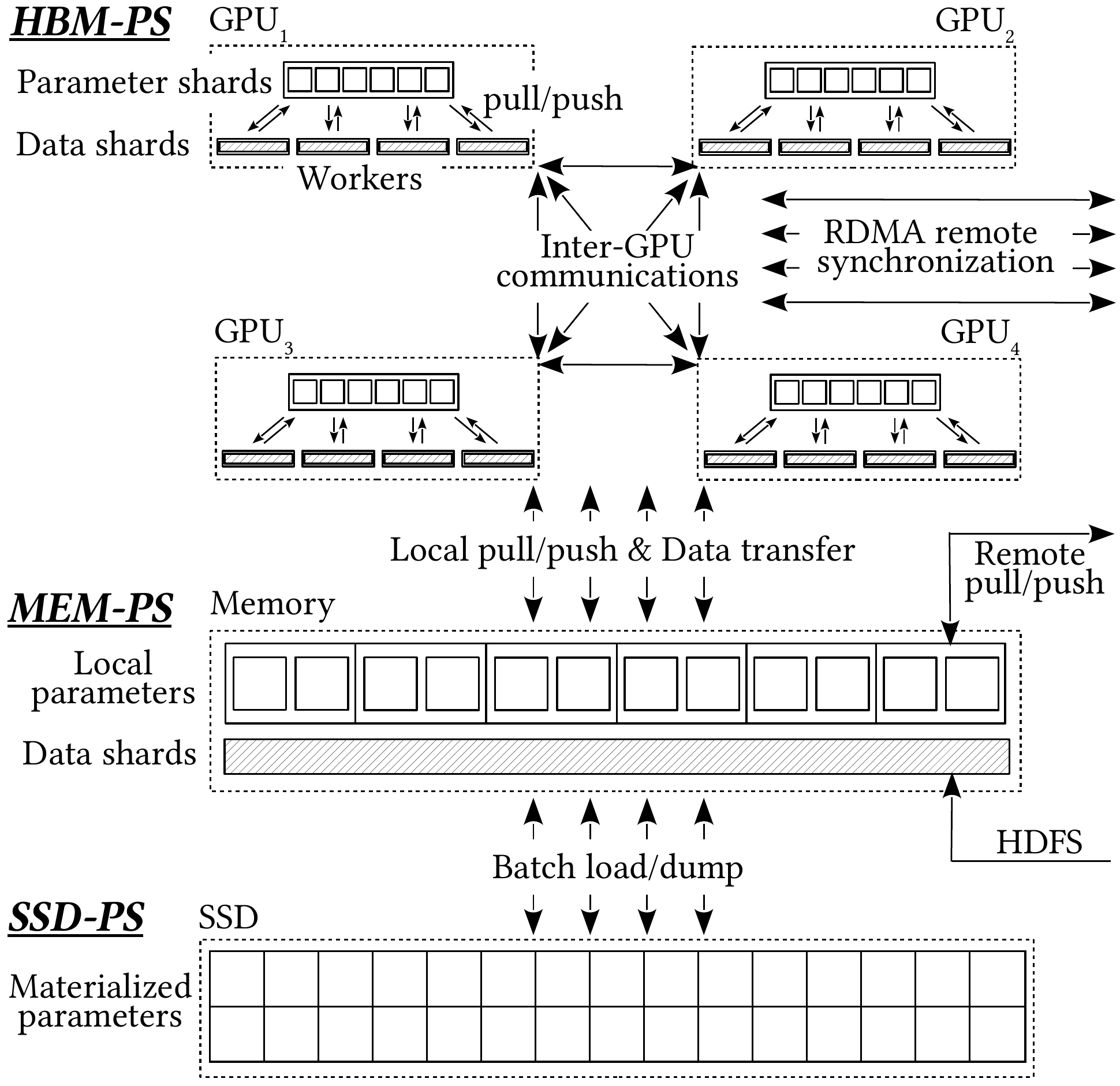}
    \vspace{-0.25in}
  	\caption{Hierarchical parameter server architecture.}
  	\label{fig:hps}
    \vspace{-0.15in}
\end{figure}

In this section, we present the distributed hierarchical parameter server overview and describe its main modules from a high-level view.
Figure~\ref{fig:hps} illustrates the proposed hierarchical parameter server architecture. It contains three major~components: HBM-PS, MEM-PS and SSD-PS.

\begin{algorithm}[h]
\caption{Distributed Hierarchical Parameter Server Training Workflow.}
\label{alg:workflow}
\algsetup{linenodelimiter=.}


\begin{algorithmic}[1]
\WHILE{\textit{not converged}}
	\STATE \textit{batch} $\leftarrow$ \textit{get\_batch\_from\_HDFS}() \label{line:hdfs}
	\STATE \textit{working}$\leftarrow$\textit{pull\_local\_MEM-PS\_and\_SSD-PS}(\textit{batch})~\label{line:pull-local}
	\STATE \textit{working} $\leftarrow$ \textit{working} $\cup$ \textit{pull\_remote\_MEM-PS}(\textit{batch})\hspace{-1cm}~\label{line:pull-remote}
	\STATE \textit{minibatches} $\leftarrow$ \textit{shard\_batch}(\textit{batch,\#GPU,\#minibatch})\hspace*{-3cm}\label{line:part-transfer-begin}
	\STATE \textit{partitions} $\leftarrow$ {\small{\textit{map\_parameters\_to\_GPUs}}}(\textit{working},\#GPU)\hspace*{-3cm}
	\FOR{$i \leftarrow1$ \textbf{to} \textit{\#GPU}}
		\STATE \textit{transfer\_to\_GPU}($i$,$\textit{minibatches}_{i}$)
		\STATE \textit{insert\_into\_hashtable}($i$,$\textit{partitions}_{i}$)
	\ENDFOR \label{line:part-transfer-end}
	\FOR{$j \leftarrow 1$ \textbf{to} \textit{\#minibatch}}
		\STATE \textit{pull\_HBM-PS\_kernel}($\textit{minibatch}_{j}$) \label{line:gpu-pull}
		\STATE 	$\Delta_{j}$ $\leftarrow$ \textit{train\_mini-batch\_kernel}($j$) \label{line:train}
		\STATE 	\textit{push\_parameters\_updates\_back\_kernel}($\Delta_{j}$) \label{line:gpu-push}
	\ENDFOR
	\STATE $\Delta$ $\leftarrow$ \textit{pull\_updates\_HBM-PS}() \label{line:hbm-collect}
	\STATE \textit{parameters\_to\_dump} $\leftarrow$ \textit{update\_local\_cache}() \label{line:dump-begin}
	\STATE \textit{push\_local\_SSD-PS}(\textit{parameters\_to\_dump}) \label{line:dump-end}
\ENDWHILE
\end{algorithmic}
\end{algorithm}

\textbf{Workflow.}
Algorithm~\ref{alg:workflow} depicts the distributed hierarchical parameter server training workflow. The training data batches are streamed into the main memory through a network file system, e.g., HDFS (line~\ref{line:hdfs}). Our distributed training framework falls in the data-parallel paradigm~\cite{li2014scaling,cui2014exploiting,Geeps,luo2018parameter}. Each node is responsible to process its own training batches---different nodes receive different training data from HDFS.
Then, each node identifies the union of the referenced parameters in the current received batch and pulls these parameters from the local MEM-PS/SSD-PS (line~\ref{line:pull-local}) and the remote MEM-PS (line~\ref{line:pull-remote}).
The local MEM-PS loads the local parameters stored on local SSD-PS into the memory and requests other nodes for the remote parameters through the network.
After all the referenced parameters are loaded in the memory, these parameters are partitioned and transferred to the HBM-PS in GPUs. In order to effectively utilize the limited GPU memory, the parameters are partitioned in a non-overlapped fashion---one parameter is stored only in one GPU. When a worker thread in a GPU requires the parameter on another GPU, it directly fetches the parameter from the remote GPU and pushes the updates back to the remote GPU through high-speed inter-GPU hardware connection NVLink~\cite{foley2017ultra}.
In addition, the data batch is sharded into multiple mini-batches and sent to each GPU worker thread (line~\ref{line:part-transfer-begin}-\ref{line:part-transfer-end}).
Many recent machine learning system studies~\cite{ho2013more,chilimbi2014project,Geeps,alistarh2018convergence} suggest that the parameter staleness shared among workers in data-parallel systems leads to slower convergence.
In our proposed system, a mini-batch contains thousands of examples.
One GPU worker thread is responsible to process a few thousand mini-batches.
An inter-GPU parameter synchronization is performed after each mini-batch is processed to eliminate the parameter staleness.
Each training worker pulls the required parameters of its corresponding mini-batch from the HBM-PS (line~\ref{line:gpu-pull}), performs forward and backward propagation to update the parameters (line~\ref{line:train}), and interacts with the HBM-PS to update the referenced parameters on other GPUs (line~\ref{line:gpu-push}).
The MEM-PS collects the updates from the HBM-PS (line~\ref{line:hbm-collect})
and dumps infrequently used parameters to the SSD-PS when the MEM-PS does not have sufficient memory (line~\ref{line:dump-begin}-\ref{line:dump-end}).
An example for Algorithm~\ref{alg:workflow} is shown in Appendix~\ref{sec:appendix:workflow-example}. HBM-PS, MEM-PS, and SSD-PS communicate in a hierarchical storage fashion. The upper-level module acts as a high-speed cache of the lower-level module. 

\textbf{HBM-PS.}
HBM-PS is distributed in the High-Bandwidth Memory (HBM) across multiple GPUs.
Comparing with the conventional distributed parameter servers, workers in GPUs can directly request and update parameters in the HBM without transferring data between GPU memory and CPU memory. Thus the training throughput is significantly increased.
However, the HBM capacity is limited and pricey.
In order to fit all the terabyte-scale parameters into HBMs, we have to maintain a distributed computing cluster with hundreds of GPUs---it is not only expensive but also inefficient, because excessive inter-GPU and inter-node communications are required to request and update the parameters. Therefore, our proposed hierarchical architecture leverages memory and SSD to store the massive model parameters.

\textbf{MEM-PS.}
The MEM-PS pulls the referenced parameters from remote nodes to prepare the data for the training in GPUs. After the GPU workers complete the forward/backward propagation, the MEM-PS retrieves the updates from the GPUs, applies the changes, and materializes the updated parameters in the SSDs.
The MEM-PS also caches the frequently used parameters to reduce SSD I/Os.

\textbf{SSD-PS.}
We build the SSD-PS to store the materialized parameters and provide efficient parameter loading and updating.
Contiguous addresses readings and writings on SSDs have better I/O bandwidth than random disk accesses. Thus, the SSD-PS organizes the parameters in files.
The updated parameters in the memory are written to SSDs in batches as new files. A file compaction thread runs in the background to regularly merge files that contain a large proportion of stale parameters into new files.

\textbf{4-stage pipeline.}
The training workflow majorly involves 4 time-consuming tasks: data transferring, parameter partitioning, materialized data loading/dumping and neural network training. The 4 tasks correspond to independent hardware resources: network, CPU, SSD and GPU, respectively.
We build a 4-stage pipeline to hide the latency of those tasks by maintaining a prefetch queue for each stage.
A worker thread is created for each stage---it extracts jobs from the prefetch queue and feeds the corresponding hardware resource. 
After that, the worker thread pushes the processed results into the prefetch queue of the next stage. Especially, the worker thread stalls when the prefetch queue of the next stage is full---the next stage has already obtained too many unprocessed jobs. The capacity of the prefetch queue is preset according to the execution time of each stage. A detailed explanation is shown in Appendix~\ref{sec:appendix:pipeline}.

\section{HBM-PS}\label{sec:hbm-ps}
The HBM-PS stores the working parameters in GPUs and provides efficient accesses. In this section, we introduce the multi-GPU distributed hash table and GPU communications. Additional details of HBM-PS are described in Appendix~\ref{sec:appendix:hbm-ps}.

\subsection{Multi-GPU Distributed Hash Table}\label{ssec:gpu-hash}
Multi-GPU distributed hash table manages the GPU HBMs on the same node and provides a unified interface to the GPU worker threads. All referenced parameters of the current training batch are partitioned and inserted into the local hash table of each GPU. Operations such as \textit{insert}, \textit{get} and \textit{accumulate} are provided to interact with the hash table.

\textbf{Local GPU hash table.}
Each GPU maintains its own local hash table in the HBM.
In this paper, we adopts the \texttt{concurrent\_unordered\_map} of cuDF library (\url{https://github.com/rapidsai/cudf}) as the hash table. The hash table implementation uses the open-addressing method to resolve hash collision.
In order to avoid the inefficient dynamic memory allocation in GPUs, we fix the hash table capacity when we construct the hash table---we know the number of parameters (key-value pairs) to be stored on each GPU. For parallel hash table updates, the data integrity is guaranteed by GPU atomic operations.


\begin{algorithm}[h]
\caption{Distributed GPU hash table accumulate.}
\label{alg:accum}
\algsetup{linenodelimiter=.}

\textbf{Input:} A collections of key-value pairs: (\textit{keys}, \textit{vals}).

\begin{algorithmic}[1]
\STATE \textit{switch\_to\_GPU}(\textit{get\_resource\_owner}(\textit{keys}, \textit{vals}))\label{line:switch-gpu-own}
\STATE \textit{partitioned} $\leftarrow$ \textit{parallel\_partition}(\textit{keys}, \textit{vals})\label{line:accum-partition}
\FOR{$i \leftarrow 1$ \textit{to} $\#GPU$} \label{line:accum-send-begin}
	\IF{$\textit{partitioned}_{i} \not= \emptyset$ }
		\STATE \textit{async\_send}($i$, $\textit{partitioned}_{i}$)
	\ENDIF
\ENDFOR \label{line:accum-send-end}
\STATE \textit{wait\_send\_destination}() \label{line:accum-wait}
\FOR{$i \leftarrow 1$ \textit{to} $\#GPU$} \label{line:accum-actual-accum-begin}
	\STATE \textit{switch\_to\_GPU}($i$)
	\STATE $\textit{hash\_tables}_{i}$.\textit{async\_accum}($\textit{partitioned}_{i}$)
\ENDFOR \label{line:accum-actual-accum-end}
\end{algorithmic}

\end{algorithm}

\textbf{Operations.}
The multi-GPU distributed hash table provides operations such as \texttt{insert}, \texttt{get} and \texttt{accumulate}.
Here we introduce the \texttt{accumulate} operation that updates the parameters during neural network training in detail.
The \texttt{insert} and \texttt{get} operations implement the pull and push interfaces to interact with the HBM-PS. Their workflows are similar to the \textit{accumulate}---their implementation details are omitted due to the page limit.

The \texttt{accumulate} operation takes a collection of key-value pairs as input and accumulates the values onto the parameters referenced by the corresponding keys.
After each back propagation is finished in the neural network training, \textit{accumulate} operations are called to update the parameters.

Algorithm~\ref{alg:accum} illustrates the \textit{accumulate} operation workflow.
The key-value pairs are passed in a zero-copy fashion---these pairs are represented as the addresses of two arrays--\textit{keys} and \textit{vals}--in the GPU HBM.
We first switch to the GPU that owns the memory of the input key-value pairs (line~\ref{line:switch-gpu-own}). Then we parallelly partition them according to the partition policy. The partitioned results for each GPU are stored at \textit{partitioned} (line~\ref{line:accum-partition}). After that, we send the partitioned key-value pairs to their corresponding GPUs (line~\ref{line:accum-send-begin}-\ref{line:accum-send-end}). The send operation is executed asynchronously through NVLink---we can send data to other GPUs simultaneously. 
Finally, we iterate over all GPUs and apply the actual accumulation on the local hash table of each GPU asynchronously (line~\ref{line:accum-actual-accum-begin}-\ref{line:accum-actual-accum-end}).

\subsection{GPU Communication}
Since each node processes its own input training batch, these batches may share some working parameters---the shared parameters are referenced and updated by multiple nodes.
The multi-GPU distributed hash table stores the working parameters across all GPUs in the same node. 
We have to synchronize the parameters on different nodes to
guarantee the model convergence~\cite{ho2013more,chilimbi2014project,Geeps,alistarh2018convergence}.

Physically, GPU communication is performed through Remote Direct Memory Access (RDMA)~\cite{potluri2013efficient} hardware.
RDMA enables zero-copy network communication---it allows the network card to transfer data from a device memory directly to another device memory without copying data between the device memory and the data buffers in the operating system. 
Our RDMA hardware design eliminates the involvement of the CPU and memory.

Logically, the inter-node GPU parameter synchronization requires an all-reduce communication---each GPU needs to receive all parameter updates from other GPUs and then performs a reduction to accumulate these updates. Appendix~\ref{sec:appendix:inter-comm} illustrates the implementation details.

\vspace{-0.08in}
\section{MEM-PS}\label{sec:mem-ps}
The MEM-PS identifies the referenced parameters from the input and communicates with the local SSD-PS and remote MEM-PS to gather the required parameters. Meanwhile, a parameter cache is maintained to reduce SSD I/Os. More implementation details are in Appendix~\ref{sec:appendix:mem-ps}.

\textbf{Prepare parameters.}
As mentioned in the distributed architecture design (Figure~\ref{fig:hps}), each node only maintains a shard of parameters---parameters are partitioned according to a pre-defined parameter-to-node mapping. We leverage the modulo hashing as the partition scheme in a similar manner to the GPU parameter partition policy discussed in Section~\ref{ssec:gpu-hash}.
The MEM-PS partitions the referenced parameters according to their keys.
For the remote parameters that belong to a remote node, the MEM-PS pulls them from other MEM-PS through the network.
For the local parameters, the MEM-PS reads the SSDs to fetch the parameters.
In order to avoid excessive SSD I/Os, an in-memory cache is maintained to store the recently/frequently used parameters and to buffer the working parameters.


\textbf{Update parameters.}
The MEM-PS pulls the updated parameters from the HBM-PS after its GPUs complete the current batch and applies these changes in the memory (we pin the working parameters in the memory).
Note that we do not need to push the changes of the remote parameters to other MEM-PS, because the remote parameters are synchronized across GPUs in HBM-PS---the remote node MEM-PS can pull the updates from its own GPUs.

\vspace{-0.05in}
\section{SSD-PS}\label{sec:ssd-ps}
The SSD-PS aims to maintain the materialized out-of-main-memory parameters efficiently on SSDs.

\textbf{Parameter files.}
We organize the model parameters in the file-level granularity---each file contains a collection of parameters. A parameter-to-file mapping is maintained in the main memory. A file descriptor occupies much less memory than the parameter value---a file descriptor can be represented as couples of bytes while a parameter value consists of multiple numeric attributes. In addition, a node only contains a shard of the parameters. Thus, we can assume the mapping can fit in the memory of a single node.

\textbf{Parameter I/O}.
We consider a parameter file as an SSD I/O unit. The SSD-PS gathers requested parameter keys and reads an entire parameter file when it contains requested parameters. On the other hand, parameters evicted from the HBM-PS cache are dumped onto SSDs. 
It is impractical to locate these parameters and perform in-place updates inside the original file because it randomly writes the disk.
Instead, our SSD-PS chunks these updated parameters into files and writes them as new files on SSDs---data are sequentially written onto the disk. After the files are written, we update the parameter-to-file mapping of these parameters.
The older versions of the parameters stored in the previous files become stale---these older values will not be used since the mapping is updated.
The SSD usage hikes as we keep creating files on SSDs.
A file compaction operation is performed regularly to reduce the disk usage---many old files containing a large proportion of stale values can be merged into new files. 
Appendix~\ref{sec:appendix:ssd-ps} presents the implementation of parameter loading/dumping and file compaction operations.

\section{Experimental Evaluation}\label{sec:experiments}
The objective of the experimental evaluation is to investigate the overall performance--as well as the impact of optimizations--of the proposed system. Specifically, the experiments are targeted to answer the following questions:
\begin{itemize}[topsep=1pt,itemsep=1pt,partopsep=1pt, parsep=1pt, leftmargin=*]
	\item How does the proposed hierarchical GPU parameter server compare with the MPI cluster solution?
	\item How is the execution time of each training stage?
	\item How does the time distribute in each component?
	\item How does the proposed system scale?
\end{itemize}



\textbf{System.}
We execute the distributed hierarchical GPU parameter server experiments on 4 GPU computing nodes. Each node has 8 cutting-edge 32 GB HBM GPUs, server-grade CPUs with 48 cores (96 threads), $\sim$1 TB of memory, $\sim$20 TB RAID-0 NVMe SSDs and a 100 Gb RDMA network adaptor.
The nodes in the MPI cluster for the baseline comparison are maintained in the same data center. CPUs in the MPI cluster have similar performance specifications as the ones in the GPU computing nodes.
All nodes are inter-connected through a high-speed Ethernet switch. The hardware and maintenance cost of 1 GPU node roughly equals to the cost of 10 CPU-only MPI nodes.\vspace{0.05in}


\begin{table}\vspace{-0.1in}\small
\caption{Model specifications.}
\label{tbl:model}
\begin{tabular}{c|clllr}
\hline \hline
 &\#Non-zeros & \#Sparse & \#Dense & Size (GB) & MPI\\
\hline
A & 100	& $8 \times 10^{9}$  & $7 \times 10^{5}$ & 300 & 100\\
B & 100	& $2 \times 10^{10}$ & $2 \times 10^{4}$ & 600 & 80\\
C & 500	& $6 \times 10^{10}$ & $2 \times 10^{6}$ & 2,000 & 75\\
D & 500	& $1 \times 10^{11}$ & $4 \times 10^{6}$ & 6,000 & 150\\
E & 500	& $2 \times 10^{11}$ & $7 \times 10^{6}$ & 10,000 & 128\\
\hline \hline
\end{tabular}
\vspace{-.5cm}
\end{table}

\begin{figure*}[htbp]
\begin{center}
\subfloat[Total execution time]{\includegraphics[width=0.33\textwidth]{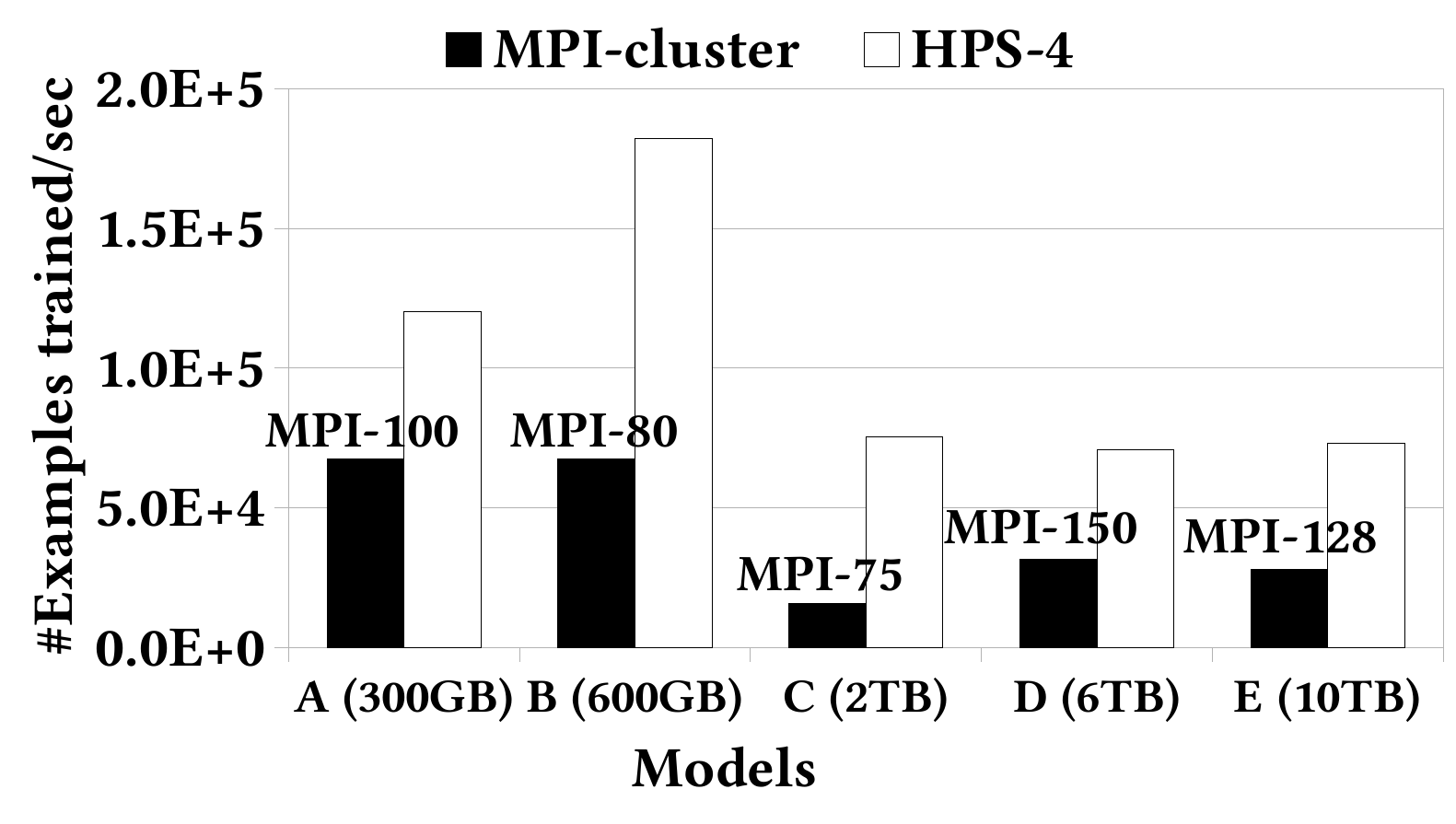}\label{fig:mpi-vs-all}}\hfill
\subfloat[AUC accuracy]{\includegraphics[width=0.33\textwidth]{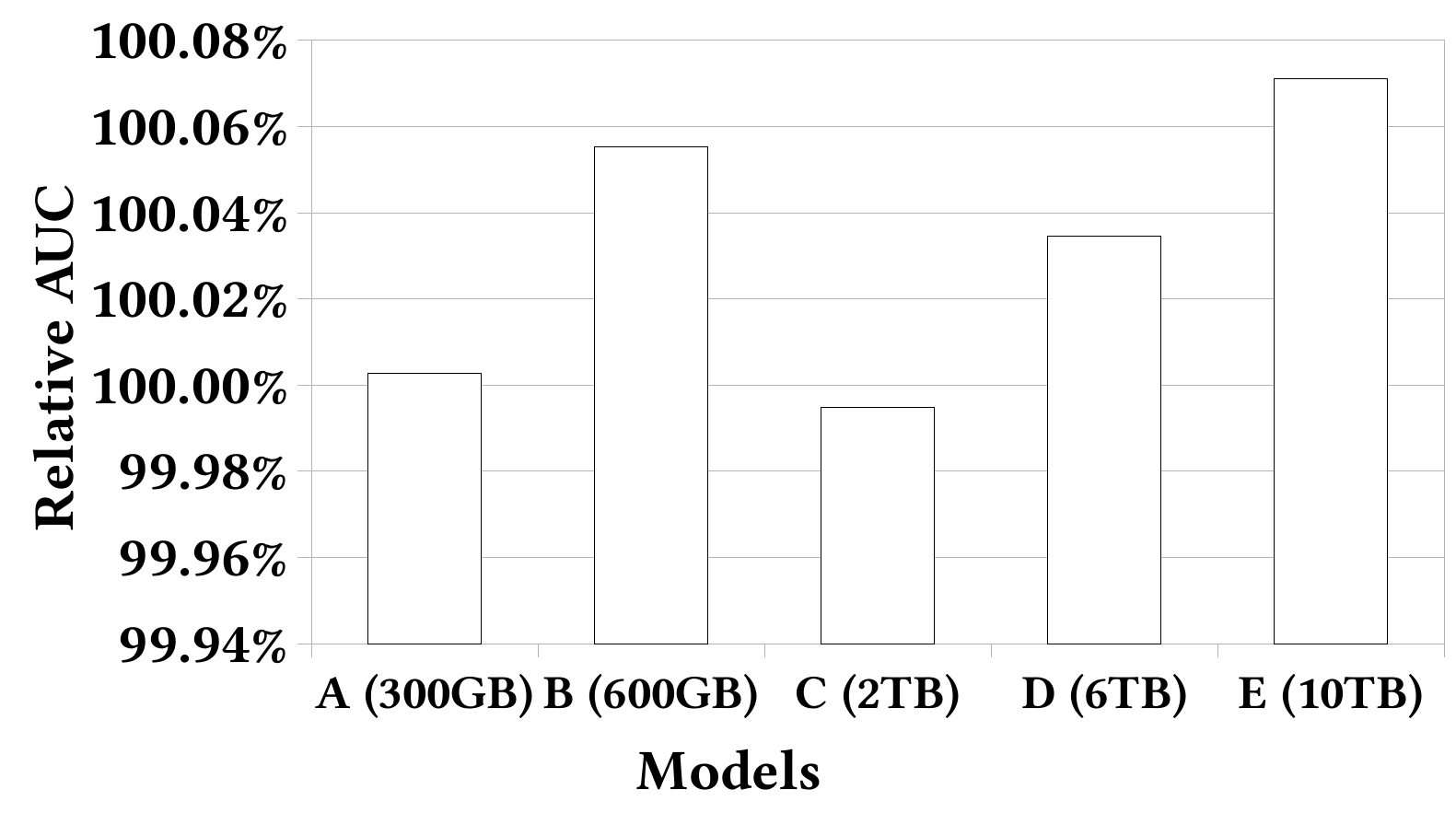}\label{fig:auc}}\hfill
\subfloat[Execution time distribution]{\includegraphics[width=0.33\textwidth]{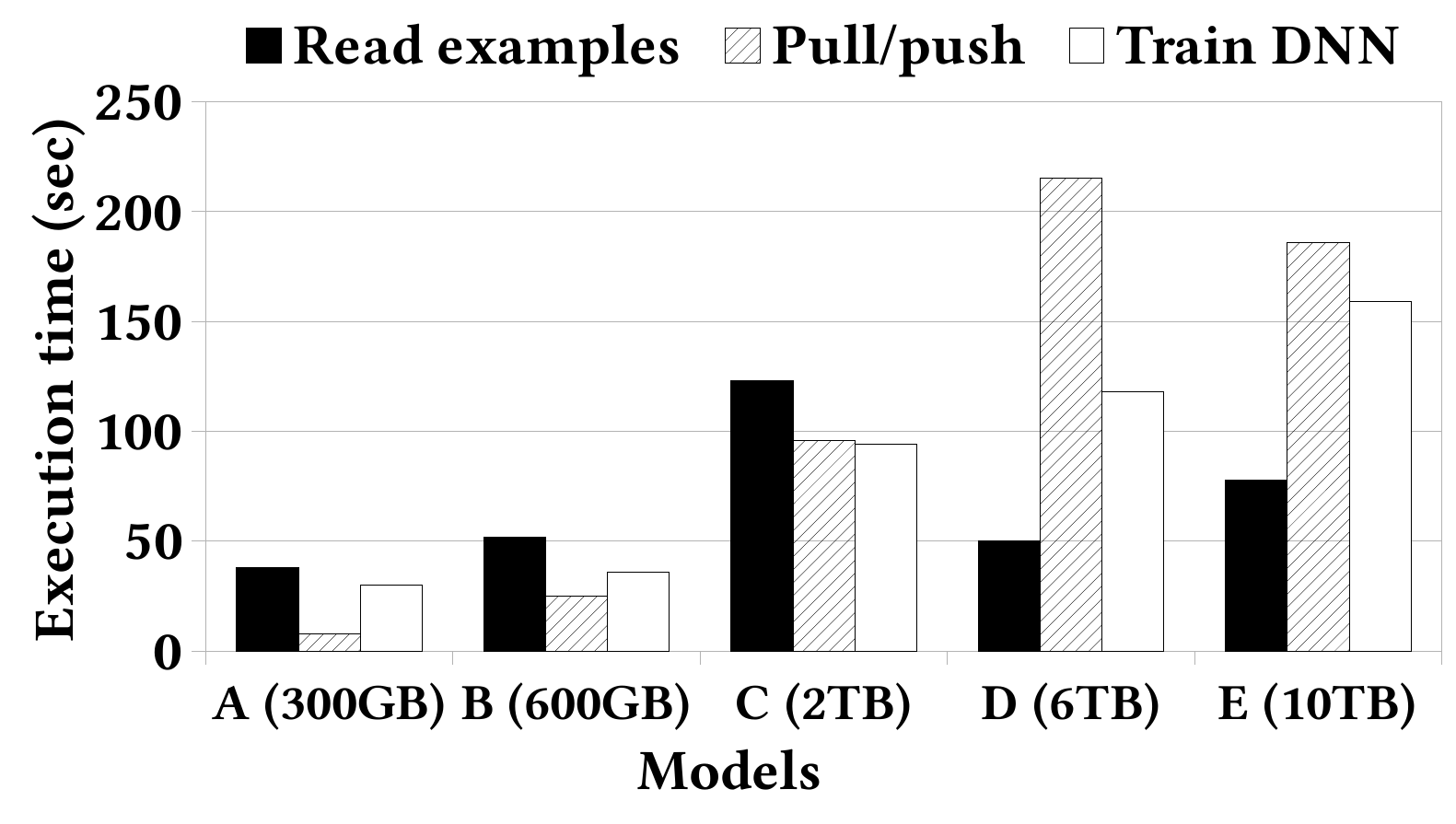}\label{fig:stage}}\hfill
\vspace{-.1in}
\caption{The performance (execution time and accuracy of 4-node hierarchical parameter server (HPS-4).}
\label{fig:mpi-auc-stage}
\end{center}
\vspace{-0.12in}
\end{figure*}

\textbf{Models.}
We use 5 CTR prediction models in real-world online sponsor advertising applications to investigate the effectiveness of our proposed system. Table~\ref{tbl:model} illustrates the specification of these models. The number of sparse parameters of each model varies from $8 \times 10^{9}$ (model A) to $10^{11}$ (model E). The number of dense parameters is $4-5$ orders of magnitude less than the number of sparse parameters for all models. The size of the smallest model is $300$ GB, while the largest model has $10$ TB parameters.
The MPI column shows the number of CPU-only nodes we used to train each model in the MPI cluster for real-world ads products. Due to different training speed requirements for these products, the number of MPI cluster nodes varies from 75 to 150.

\textbf{Data.}
We collect user click history logs from our search engine as the training dataset.
The datasets are chunked into batches---each batch contains $\sim$$4 \times 10^{6}$ examples. The trained CTR models are evaluated over the production environment of our search engine in an A/B testing manner.

\vspace{-0.05in}
\subsection{Comparison with MPI Solution}
We evaluate the performance of the proposed system on the training execution time and the prediction accuracy.
Our proposed system runs on $4$ GPU computing nodes.
We take an MPI cluster training framework in the production environment as the baseline.
The baseline solution employs a distributed parameter server that shards and keeps the massive parameters in the memory across all the nodes.

\begin{table}[b]\vspace{-0.25in}\small
\caption{Training speedup over the MPI-cluster solution.}
\label{tbl:mpi-speedup}
\centering
\begin{tabular}{r|ccccc}
\hline
\hline
 & A & B & C & D & E\\
\hline
Speedup over MPI-cluster & 1.8 & 2.7 & 4.8 & 2.2 & 2.6\\
Cost-normalized speedup & 4.4 & 5.4 & 9.0 & 8.4 & 8.3\\
\hline
\hline
\end{tabular}
\end{table}

\textbf{Training time.}
Figure~\ref{fig:mpi-auc-stage}(a) depicts the total training execution time of the $4$-node hierarchical parameter server (HPS-4) and the MPI training solution (MPI-cluster). HPS-4 outperforms MPI-cluster in all $5$ models. The first row of Table~\ref{tbl:mpi-speedup} shows the speedup over the MPI solution. The HPS-4 is 1.8-4.8X faster than the MPI solution.
Note that the hardware and maintenance cost of 1 GPU node is roughly equivalent to 10 CPU-only nodes
 in the MPI cluster. The 4-GPU-node setting is much cheaper than the 75-150 nodes in the MPI cluster.
 The second row of the table illustrates the cost-normalized speedup of our proposed system. The cost-normalized speedup is computed as: $\textit{speedup} / 4 / 10 \times \#\text{MPI}$.
 The price-performance ratio of our proposed system is 4.4-9.0X better than the MPI cluster. 


\textbf{Accuracy.}
We take Area Under the Curve~\cite{huang2005using} (AUC) as the quality measure of our trained models.
Figure~\ref{fig:mpi-auc-stage}(b) shows the relative AUC of HPS-4 (relative to the MPI-cluster method). The AUC of the MPI-cluster solution is $100\%$. The AUCs of both trained models are tested online for $1$ day in real online ads products. Since the CTR prediction accuracy is crucial to the revenue, we have to ensure all our optimizations are loss-less. For Model C, the relative AUC loss of HPS-4 is less than $0.01\%$. For other $4$ models, HPS-4 has even slightly better accuracy than the MPI-128. Our hierarchical parameter server runs on fewer nodes---fewer stale parameters are used than the case on the MPI cluster. Therefore, it is possible to have a better AUC than the MPI solution.
Overall, the relative differences of all $5$ models are within $0.1\%$. We can conclude that our hierarchical parameter server training is loss-less.

\vspace{-0.05in}
\subsection{Time distribution}
The time distribution of each abstract execution stage is presented in Figure~\ref{fig:mpi-auc-stage}(c). As illustrated in the training workflow (Algorithm~\ref{alg:workflow}), \texttt{Read examples} corresponds to the operation that reads and extracts the data from HDFS; \texttt{Pull/push} is the execution time of gathering and updating the working parameters in the MEM-PS and SSD-PS; and \texttt{Train DNN} is the execution time of the inter-GPU communication and the deep neural network training in the HBM-PS.
Since these stages are paralleled in the pipeline, the overall execution time for each batch is dominated by the slowest stage.
For the two small models (Model A and B), the \texttt{Read examples} stage that relies on the HDFS throughput becomes the bottleneck. Because the models are small and fit in the memory, the sparse parameter pulling and pushing operations are faster than the HDFS I/Os. When the number of sparse parameters grows, the \texttt{Pull/push} operations catch up with the example reading (Model C) and become the dominant stage in Model D and E.

\vspace{-0.05in}
\subsection{HBM-PS}
\begin{figure*}[t]
\begin{center}
\subfloat[Time distribution in HBM-PS]{\includegraphics[width=0.33\textwidth]{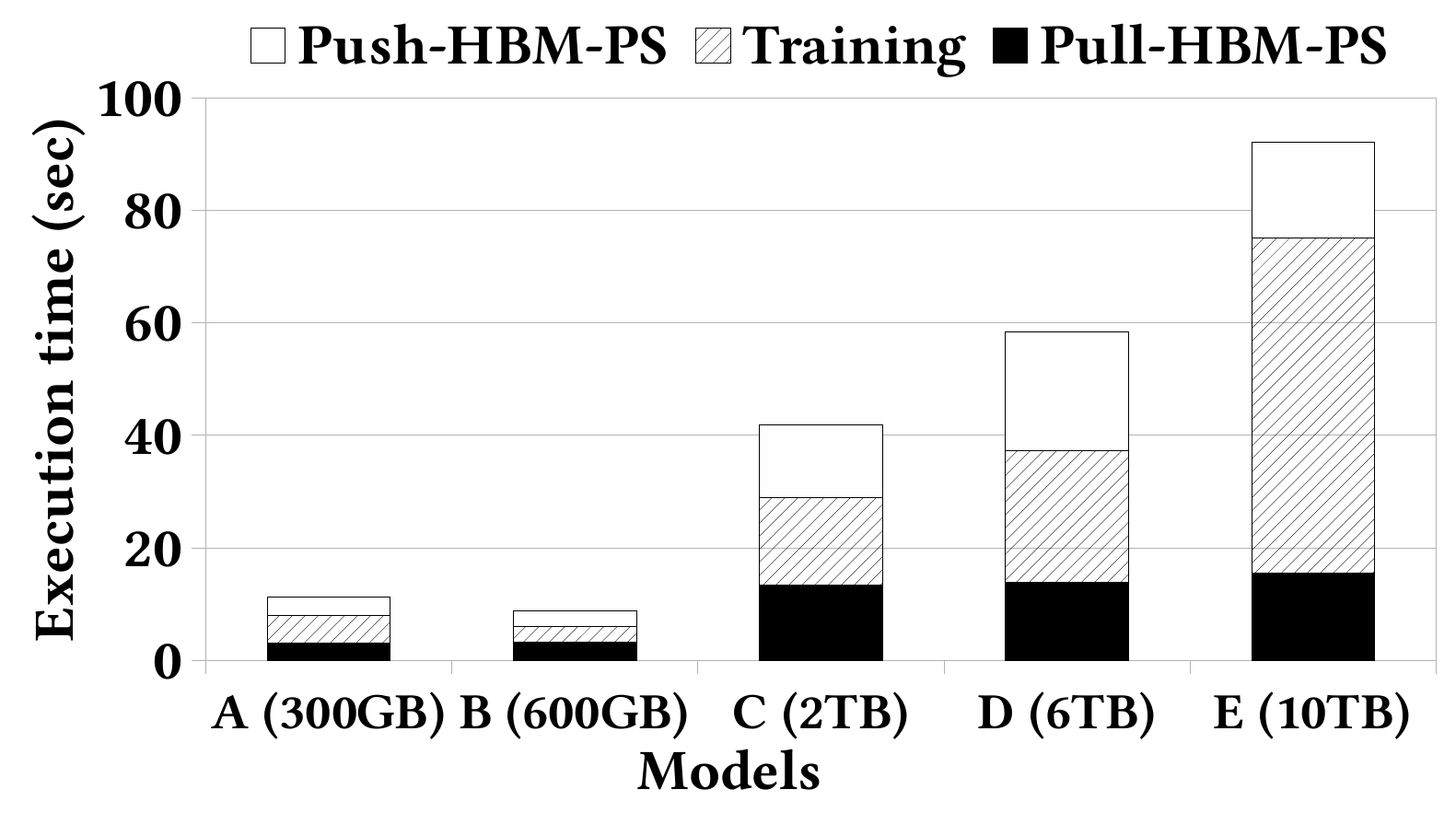}\label{fig:inside-hbm}}\hfill
\subfloat[Time distribution in MEM-PS]{\includegraphics[width=0.33\textwidth]{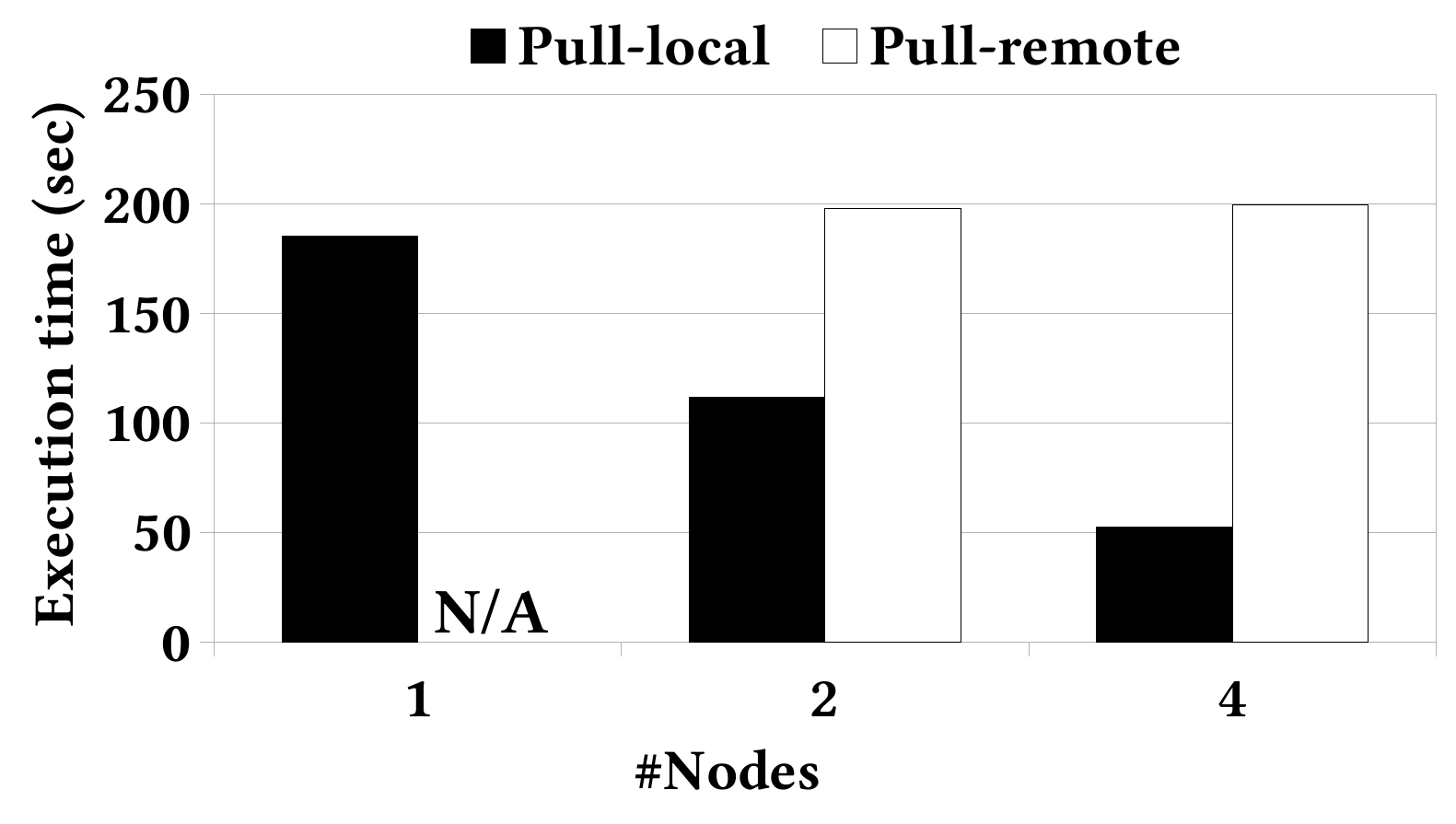}\label{fig:memps-pull}}\hfill
\subfloat[Cache hit rate]{\includegraphics[width=0.33\textwidth]{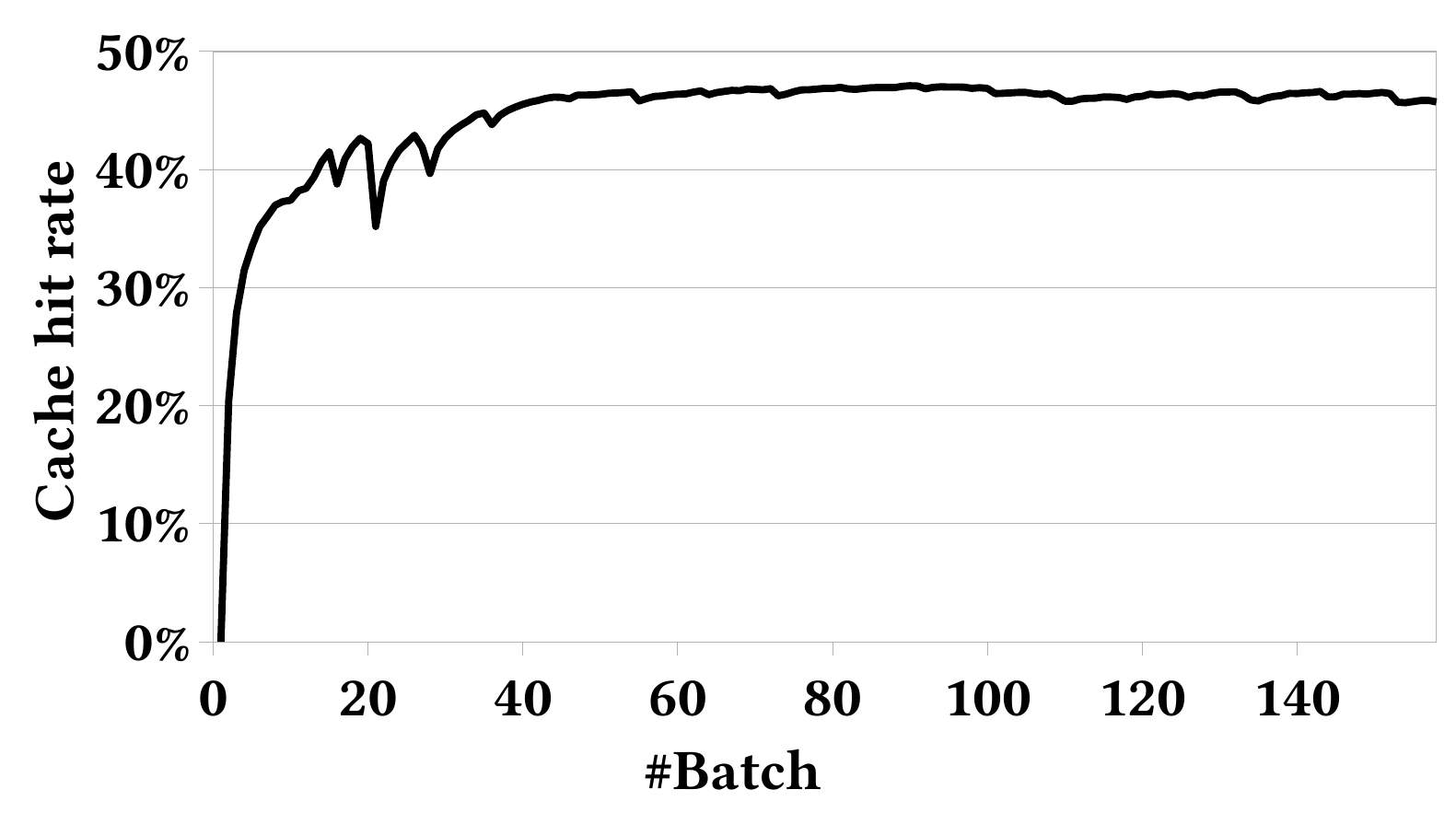}\label{fig:cache-hit}}\hfill
\vspace{-0in}
\caption{The time distribution in HBM-PS/MEM-PS and the cache hit rate on Model E.}
\label{fig:hbm-mem-cache}
\end{center}
\end{figure*}

The execution time distribution of HBM-PS operations is shown in Figure~\ref{fig:hbm-mem-cache}(a).
The pull/push operations of HBM-PS depend on the number of non-zero values of training inputs. While the training operation (forward/backward propagation) correlates to the number of dense parameters of the deep model. The execution time of these operations shows the same trend in the model specification (Table~\ref{tbl:model}). Model A and Model B have around $100$ non-zero input features per example, the pull/push HBM-PS operations are faster than the ones of Model C, D and E. Model E has the largest number of dense parameters. Thus its training operation costs most of the execution time in HBM-PS.

\subsection{MEM-PS}
\textbf{Local/remote parameters.}
Figure~\ref{fig:hbm-mem-cache}(b) illustrates the execution time of pulling local/remote parameters in the MEM-PS for Model E over 1, 2 and 4 GPU nodes.
The results of other models show a similar trend. 
The remote parameter pulling operation is not applicable when we only have $1$ GPU node---all the parameters are stored in the same node.
When we deploy the proposed system in distributed environments with 2 and 4 nodes, the local and remote pulling operations are paralleled. The overall execution time is determined by the slower operation. We can observe that the overall time of pulling MEM-PS parameters does not hike much when inter-node communications are involved. 

\vspace{0.08in}

\textbf{Cache performance.}
The cache hit rate for Model E is illustrated in Figure~\ref{fig:hbm-mem-cache}(c).
In the beginning of the training, the cache hit rate is low because we start from a cold cache---no data are cached. We can observe that the cache hit rate increases steeply during the training of the first $10$ batches. After $40$ training batches, the cache hit rate goes to $46\%$ and becomes stable in the following training---the frequently visited parameters are captured by our cache policy.

\begin{figure}[t]
\hspace{-.1in}
\subfloat[SSD-PS I/O time]{\includegraphics[width=0.26\textwidth]{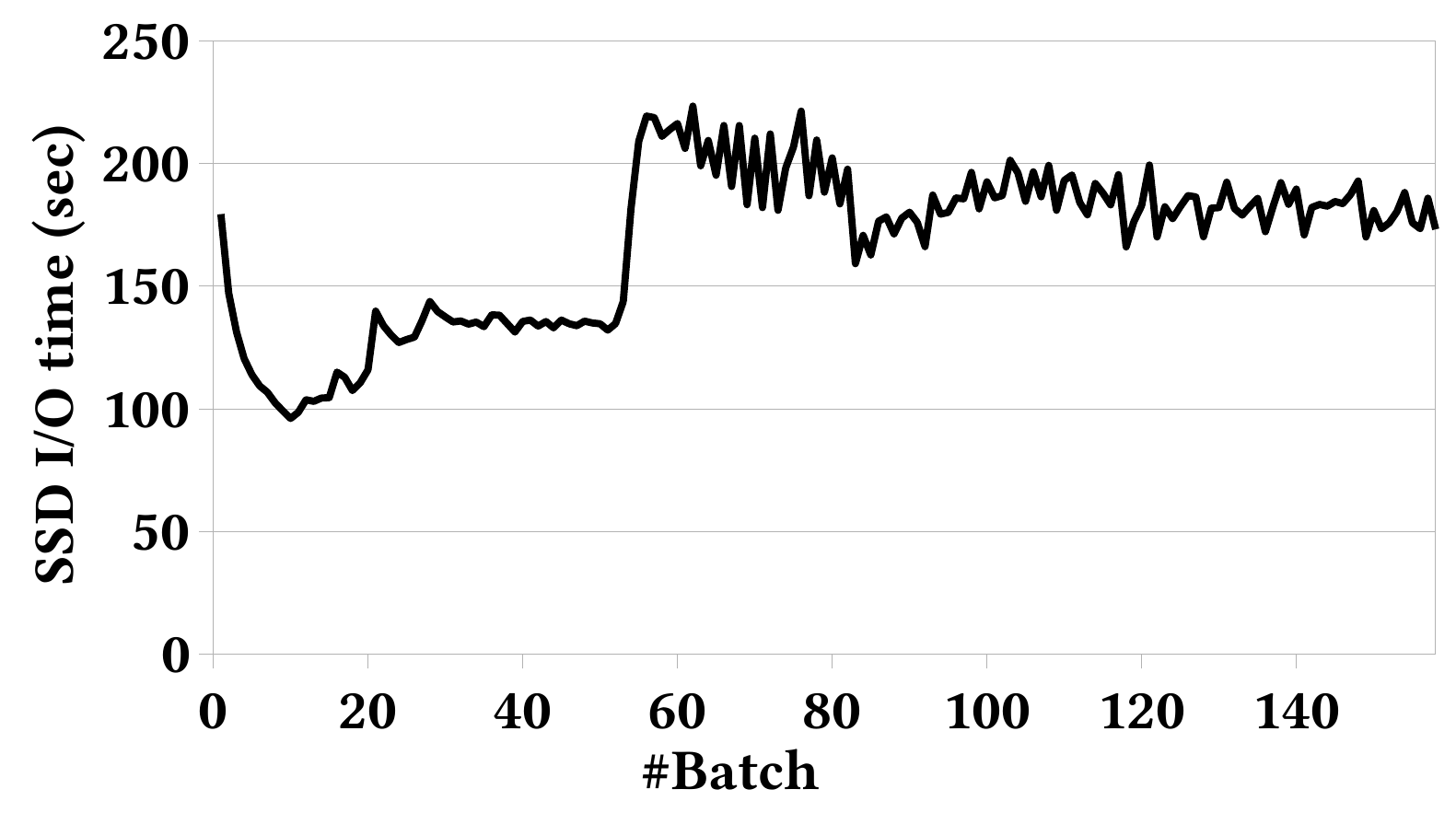}\label{fig:ssdps-io}}
\subfloat[Speedup]{\includegraphics[width=0.26\textwidth]{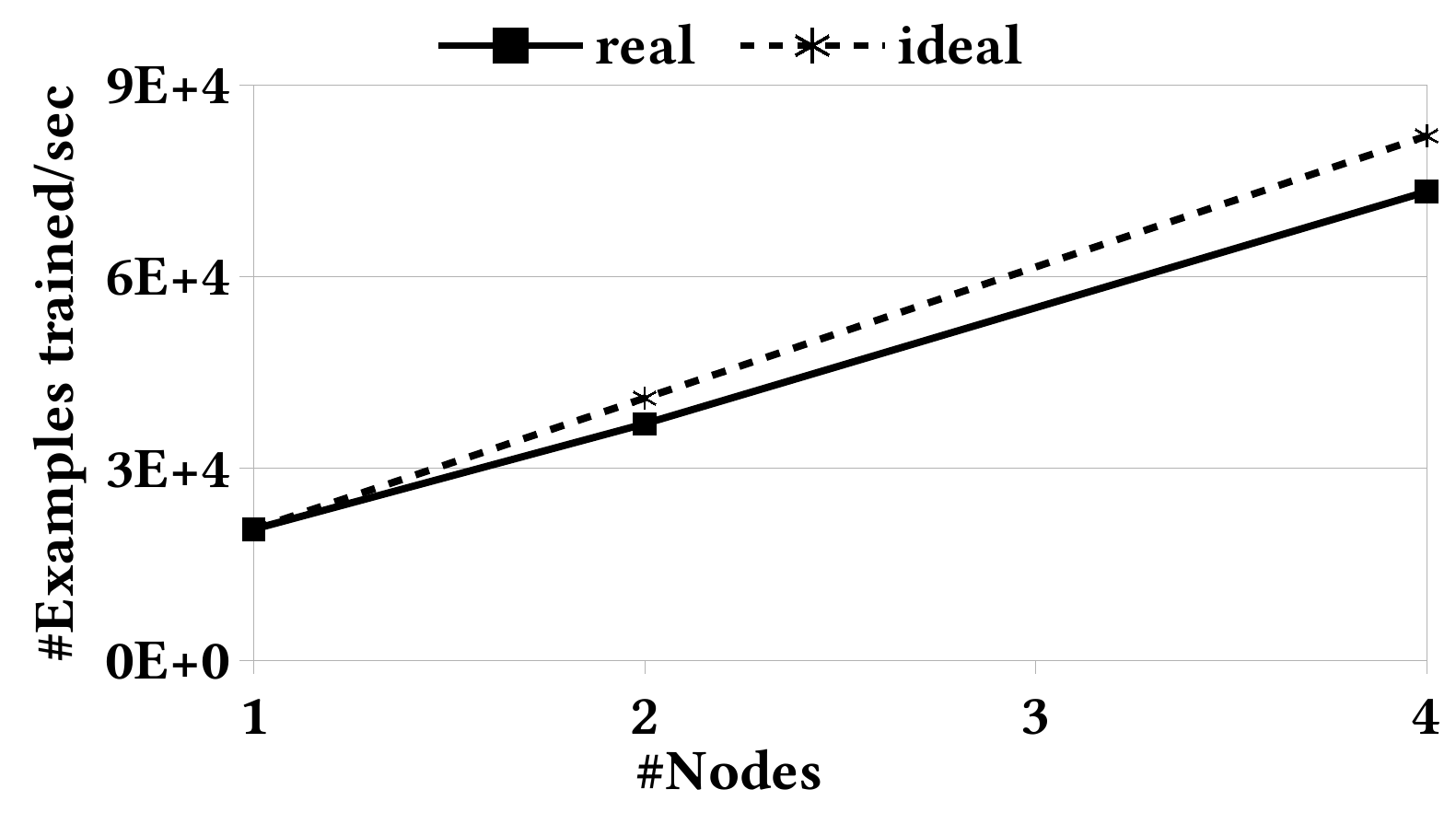}\label{fig:speedup}}
\vspace{-0in}
    \caption{The SSD-PS I/O time and  speedup on Model E.}
    \label{fig:ssd-speedup}
    \vspace{-.1in}
\end{figure}

\vspace{0.05in}
\subsection{SSD-PS}
Figure~\ref{fig:ssd-speedup}(a) presents the SSD I/O time for Model E.
The SSD disk usage threshold is reached and the parameter file compaction operation is involved from the $54^{\textit{th}}$ batch---the SSD I/O time hikes.
The regular file merging in the compaction operation causes the I/O performance fluctuation.

\subsection{Scalability}
The training throughput (number of examples trained per second) speedup is depicted in Figure~\ref{fig:ssd-speedup}(b) over 1, 2, and 4 nodes. The dashed line represents the ideal linear speedup.
The reason for the (slightly) sub-linear speedup is that more network communications are incurred when we use more nodes. A speedup of 3.57 out of 4 is obtained for 4 nodes.

\subsection{Discussion}
Based on the results shown above, we can answer the questions driving the experimental evaluation.
The 4-node distributed hierarchical GPU parameter server is 1.8-4.8X faster than the MPI solution in the production environment.
The cost of 4 GPU nodes is much less than the cost of maintaining 75-150 CPU nodes in the MPI cluster.
After normalizing the execution time by the hardware and maintenance cost, the price-performance ratio of our proposed system is 4.4-9.0X better than the MPI solution. Besides the training speed, the relative accuracy difference to the MPI solution is within 0.1\%---the training in our proposed system is loss-less.
The proposed pipeline parallels all training stages. The overall execution time for each batch is dominated by the HDFS when the model is small.
When the number of sparse parameters grows to 6 and 10 TB, the sparse parameter pulling and pushing operations become the dominant factor.
The execution time in the HBM-PS depends on the model specifications.
A larger number of non-zero values of training inputs yields longer execution time for pulling and pushing parameters in the HBM-PS.
The training forward and backward propagation time depend on the size of fully-connected layers in the deep model.
The execution time of pulling working parameters in the MEM-PS is almost independent of the number of nodes. Although the local data required to load from SSDs are reduced when we have multiple nodes, more parameter requests from remote nodes have to be processed. In expectation, the average number of parameters to load from SSDs stays the same in multi-node cases.
The SSD-PS performance has slight fluctuations when the parameter file compaction operation involves in.
Our proposed system scales well in distributed environments---a training throughput speedup of 3.57 out of 4 (the ideal speedup) is obtained for 4 nodes.

\section{Related Work}\label{sec:related}
In this section, we discuss relevant work from CTR prediction models and parameter servers. Additional related work about in-memory cache management systems and key-value stores on SSDs are discussed in Appendix~\ref{sec:appendix:relwork}.

\vspace{0.08in}

\textbf{CTR prediction models.}
The large scale logistic regression model with careful feature engineering used to dominate the CTR prediction strategies~\cite{edelman2007internet,graepel2010web}. Recently, deep neural networks with embedding layers are applied in the CTR prediction problem and obtain significant improvements. Generally, the sparse input features are converted into dense vectors through embedding layers, and then feed into neural components to expose the correlation/interaction among them.
The differences among these models lie in the neural components above the embedding layer. For example, Deep Crossing~\citep{shan2016deep}, Wide\&Deep Learning~\citep{cheng2016wide}, YouTube Recommendation CTR model~\citep{covington2016deep} and Deep Interest Network (DIN)~\citep{zhou2018deep}  design special fully-connected layers for corresponding tasks to capture the latent interactions among features; Product-based Neural Network (PNN)~\citep{qu2016product} employs a product layer to capture high-order feature interactions; DeepFM~\citep{guo2017deepfm} and xDeepFM~\citep{lian2018xdeepfm} use factorization machines (FM) to model both low-order and high-order feature correlations. Through this approach, the correlations between sparse features are automatically exploited. 

The successes of these systems have proven that deep learning with sparse input is an effective approach for real-world commercial advertisement and recommendation systems. However, when the scale of the problem becomes extremely huge, for example, with $10^{11}$ dimensional sparse data and trillions of samples, the architecture of these models must be carefully designed to cope with the challenges mentioned in above sections. This paper is our attempts to solving these challenges using an SSD based, three-layer distributed GPU training architecture. We believe our solution greatly enhances the feasibility of the methodology of deep learning with sparse input in much larger scale problems. 

\vspace{0.08in}

\textbf{Parameter servers.}
One important research direction is the parallel mechanism among the computational nodes and server nodes. One  such topic  is the synchronization pattern of local updates of parameters, where three typical types of synchronization patterns are proposed: 1) Bulk Synchronous Parallel (BSP)~\citep{valiant1990BSP}, which strictly synchronizes all local updates from all computing nodes in each iteration, and thus the learning procedure acts as the same logic as that of learning in a single machine. 2) Stale Synchronous Parallel (SSP)~\citep{ho2013more}, which allows the fastest computing node to run a certain number of iterations ahead of other nodes. 3) Asynchronous Parallel (ASP), which does not need any synchronization among the updates of computing nodes. Instead of using the same parallel strategy throughout the training procedure, FlexPS~\cite{FlexPS} proposed a flexible parallelism control for multi-stage~learning algorithms such as~SVRG. 

Another direction is to design efficient parameter server architecture for different hardware/network  situations. Communication is a major bottleneck in parallel training.  Poseidon~\citep{Poseidon} develops two strategies \emph{wait-free back-propagation} and \emph{hybrid communication} to exploit the independency between parameters of layered model structures in deep neural networks and conquer communication bottleneck; while Gaia~\cite{Gaia} even considered the problem of running learning program on geo-distributed clusters that communicate over WANs. Another challenge is the big model size for GPU platforms. When the model becomes bigger, the limited GPU memory cannot hold the entire model. GeePS~\citep{Geeps} builds a two-layer architecture to cope with the problems in this situation, including data movement overheads, GPU stalls, and limited GPU memory, where the Memory layer holds the entire model in memory, and the GPU layer fetch the required part of the model from the memory layer, do the computation and rewrite the updates to the memory layer. Our proposed system constructs a distributed hash table across GPU HBMs that enables direct GPU peer-to-peer communications. It reduces the excessive CPU-GPU data movement and synchronization overhead in GeePS.

However, when the model size becomes even larger, the parameters cannot even be stored in the CPU main memory of the cluster. Our solution to the challenge is a three-layer architecture that consists of SSD-PS, MEM-PS layer, and HBM-PS. The new architecture results in new challenges such as GPU I/O problems and we develop strategies to optimize the performance. Our architecture can efficiently train a 10 TB model, which greatly enhances the capability of GPU clusters to cope with massive scale models.

\section{Conclusions}\label{sec:conclusion}
Since 2013, Baidu
Search Ads (a.k.a. ``Phoenix Nest'') has been successfully using ultra-high dimensional input data and ultra-large-scale deep neural networks for training
CTR prediction models. In this paper, we introduce the architecture of a distributed hierarchical GPU parameter server for massive scale deep learning ads systems. As discussed in Section~\ref{sec:hash}, model hashing techniques such as ``one-permutation + one sign random projection'' are not fully applicable to the accuracy-crucial ads industry applications. The deep learning model parameter size can become huge and cannot fit in the CPU/GPU memory. 
The proposed hierarchical design employs the main memory and SSDs to, respectively, store the out-of-GPU-memory and out-of-main-memory parameters of massive scale~neural~networks.

We perform an extensive set of experiments on $5$ CTR prediction models in real-world online sponsor advertising applications. The results confirm the effectiveness and the scalability of the proposed system.
The 4-node distributed hierarchical GPU parameter server is 1.8-4.8X faster than the MPI solution in the production environment. For example, a 4-node hierarchical GPU parameter server can train a model more than 2X faster than a 150-node in-memory distributed parameter server in an MPI cluster on Model~D.
The cost of 4 GPU nodes is much less than the cost of maintaining an MPI cluster of 75-150 CPU nodes.
After normalizing the execution time by the hardware cost and maintenance expenditure, the price-performance ratio of this proposed system is 4.4-9.0X better than the previous MPI solution. 

The system described in this paper is being integrated with the PaddlePaddle deep learning platform (\url{https://www.paddlepaddle.org.cn}) to become the ``PaddleBox''. 

\vspace{-0.05in}
\section*{Acknowledgement}
\vspace{-0.05in}

The system described  in this paper represents  the substantial effort in the past years involving several major product teams at Baidu Inc. We could not include all the names we should acknowledge but only a few: Xuewu Jiao, Quanxiang~Jia, Lian~Zhao, ~Lin~Liu,~Jiajun~Zhang,~Yue~Wang,~Anlong~Qi. 
%

%
\clearpage
\balance
\bibliographystyle{mlsys2020}
\bibliography{biblio,standard}

\clearpage
\appendix

\section{Hierarchical Parameter Server Workflow Example}\label{sec:appendix:workflow-example}
\begin{figure}[tb]
	\centering
	\includegraphics[width=.45\textwidth]{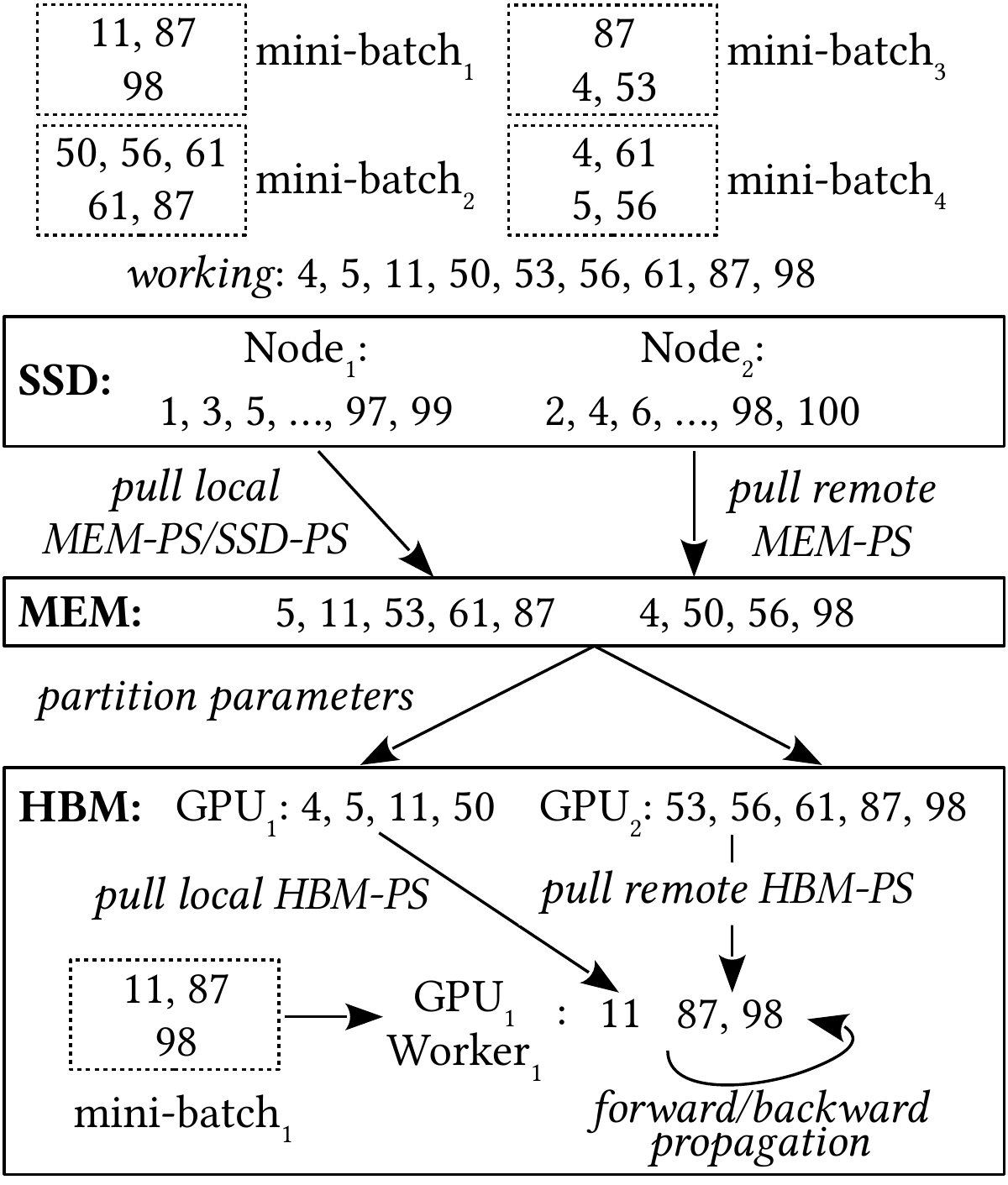}
  	\caption{An example for Algorithm~\ref{alg:workflow}.}
  	\label{fig:workflow-example}
\end{figure}

\textbf{Example.} Figure~\ref{fig:workflow-example} depicts an example for the training workflow (Algorithm~\ref{alg:workflow}). Consider now we are at $\textit{node}_{1}$. An input batch is streamed from HDFS and is divided into $4$ mini-batches. The working parameters of the current batch are: $4, 5, 11, 50,53,56, 61, 87, 98$.
Parameters are sharded and stored on the SSDs of each node.
We have $2$ nodes and shard the parameters in a round-robin method in this example---$\textit{node}_{1}$ stores the parameters with odd keys while $\textit{node}_{2}$ stores the ones with even keys. Here we have $100$ total parameters in this example---there are $10^{11}$ parameters in real-world large-scale deep learning models. $5, 11, 53, 61, 87$ are stored on the local node---$\textit{node}_{1}$. We pull these parameters from the local MEM-PS (for the cached parameters) and the local SSD-PS. For the parameters that stored on other nodes---$4, 50, 56, 98$, we pull these parameters from the MEM-PS on $\textit{node}_{2}$ through the network. The MEM-PS on $\textit{node}_{2}$ interacts with its memory cache and its local SSD-PS to load the requested parameters. Now all the working parameters are retrieved and are stored in the memory of $\textit{node}_{1}$. Here we have $2$ GPUs on $\textit{node}_{1}$. The working parameters are partitioned and transferred to GPU HBMs. In this example, $\textit{GPU}_{1}$ obtains the parameters whose keys are less than or equal to $50$---$4, 5, 11, 50$, and $\textit{GPU}_{2}$ takes $53, 56, 61, 87, 98$. The partition strategy can be any hashing function that maps a parameter key to a GPU id. Consider the $\textit{worker}_{1}$ of $GPU_{1}$ is responsible to process $\textit{mini-batch}_{1}$. $\textit{worker}_{1}$ is required to load $11$, $87$ and $98$. Among them, $11$ is stored in the HBM of the local GPU---$\textit{GPU}_{1}$. $87$ and $98$ are pulled from $\textit{GPU}_{2}$. Since the GPUs are connected with high-speed interconnections---NVLink, the inter-GPU data transfer has low-latency and high-bandwidth. After the parameters are ready in the working memory of $\textit{worker}_{1}$, we can perform the neural network forward and backward propagation operations to update the parameters. All the updated parameters are synchronized among all GPUs on all nodes after each mini-batch is finished. When all the mini-batches are finished, the MEM-PS on each node pulls back the updated parameters and materializes them onto~SSDs.

\section{4-stage Pipeline Example}\label{sec:appendix:pipeline}
\begin{figure}[htbp]
	\hspace*{-.3in}	
	\includegraphics[width=.54\textwidth]{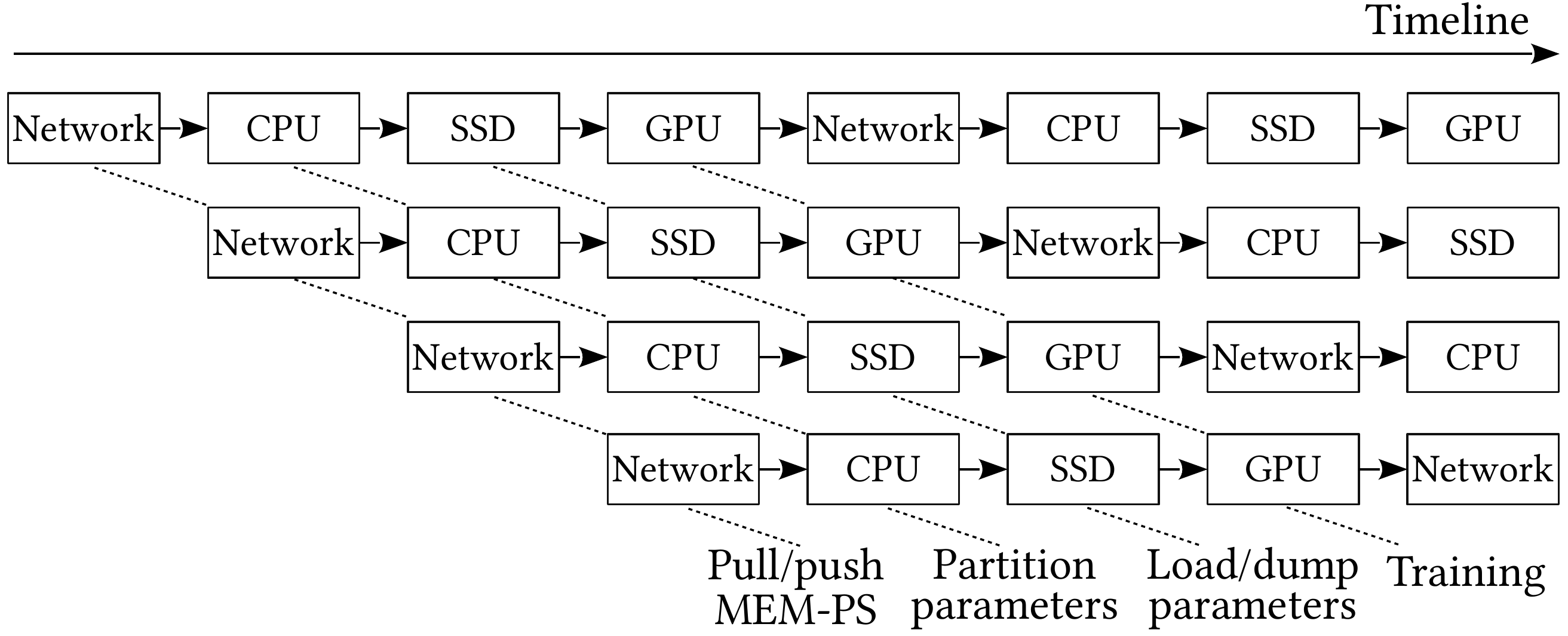}
  	\caption{The 4-stage pipeline.}
  	\label{fig:pipeline}
\end{figure}
Figure~\ref{fig:pipeline} is an illustration of the 4-stage pipeline.
For example, when the GPUs are busy training the model, our 4-stage pipeline enables the proposed system to prepare the referenced parameters of the next training batch at the same time: the HBM-PS pulls remote parameters from other nodes and the SSD-PS loads local parameters for the next batch simultaneously. After the training of the current batch is finished, all the required parameters of the next batch are ready to use in the GPU HBM---GPUs are able to train the next batch immediately.

\begin{figure*}[htbp]
\centering
	\includegraphics[width=.9\textwidth]{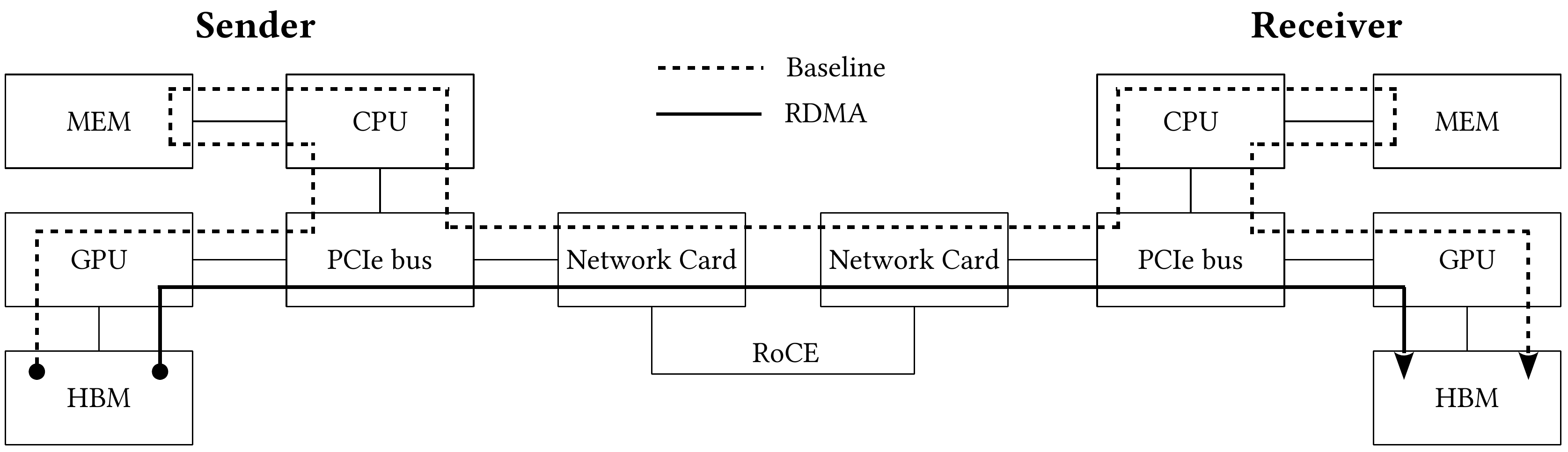}
 	\caption{Inter-node RDMA communication.}
  	\label{fig:rdma}
\end{figure*}

\section{HBM-PS Implementation}\label{sec:appendix:hbm-ps}
\subsection{Multi-GPU Distributed Hash Table}\label{ssec:appendix:gpu-hash}
\textbf{Partition policy.}
A partition policy that maps a parameter key to a GPU id is required to partition the parameters. A simple modulo hash function yields a balanced partitioning in general cases, because the features of the input training data are usually distributed randomly.
The modulo hash function can be computed efficiently with constant memory space. 
As a trade-off of the memory footprint, the disadvantage of the simple hash partition policy is that we need to pull parameters from other GPUs if the parameters referenced in a mini-batch are not stored in the local parameter partition.
One possible improvement is to group parameters with high co-occurrence together~\cite{eisenman2018bandana}, for example, pre-train a learned hash function~\cite{kraska2018case} to maximize the parameter co-occurrence. It is another research axis--vertical partitioning~\cite{navathe1984vertical,zhao2015vertical} that is beyond the scope of this paper.
Generally, no perfect balanced partition solution exists for random inputs. Although the number of pulled parameters is reduced, we still have to pull parameters from almost all other GPUs even with an optimized partition policy.
Besides, transferring a large batch of data can better utilize the NVLink bandwidth---the disadvantage of the simple hash function partition policy is reduced.

\subsection{GPU RDMA Communication}\label{sec:appendix:rdma}
\textbf{GPU Communication mechanism.}
The inter-node GPU communication mechanism is depicted in Figure~\ref{fig:rdma}. Two nodes are shown in the figure---the left node is the sender and the right one is the receiver. For each node, the CPU, GPU and network card are connected with a PCIe bus.

We first examine the \textit{baseline} method before we introduce our \textit{RDMA} solution. In the figure, the data flow of the baseline method is represented as the dashed line starting from the sender HBM to the receiver HBM. The CPU calls the GPU driver to copy the data from GPU HBM into the CPU memory. Then, the CPU reads the data in the memory and transmits the data through the network. The transmitted data are stored in the memory of the receiver node. Finally, the receiver CPU transfers the in-memory data to the GPU HBM.
In this baseline method, the CPU memory is utilized as a buffer to store the communication data---it incurs unnecessary data copies and CPU consumptions.

Our \textit{RDMA} hardware design eliminates the involvement of the CPU and memory. Its data flow is represented as a solid line in the figure.
Remote Direct Memory Access (RDMA)~\cite{potluri2013efficient} enables zero-copy network communication---it allows the network card to transfer data from a device memory directly to another device memory without copying data between the device memory and the data buffers in the operating system. 
The RDMA data transfer demands no CPU consumption or context switches.
The network protocol--RDMA over Converged Ethernet (RoCE)--is employed to allow RDMA over the Ethernet network. The sender GPU driver\footnote{\url{https://docs.nvidia.com/cuda/gpudirect-rdma/index.html}} directly streams the data in HBM to the network, while the receiver network card directly stores the collected data into the GPU HBM.

\begin{figure}[htbp]
	\centering
	\includegraphics[width=.5\textwidth]{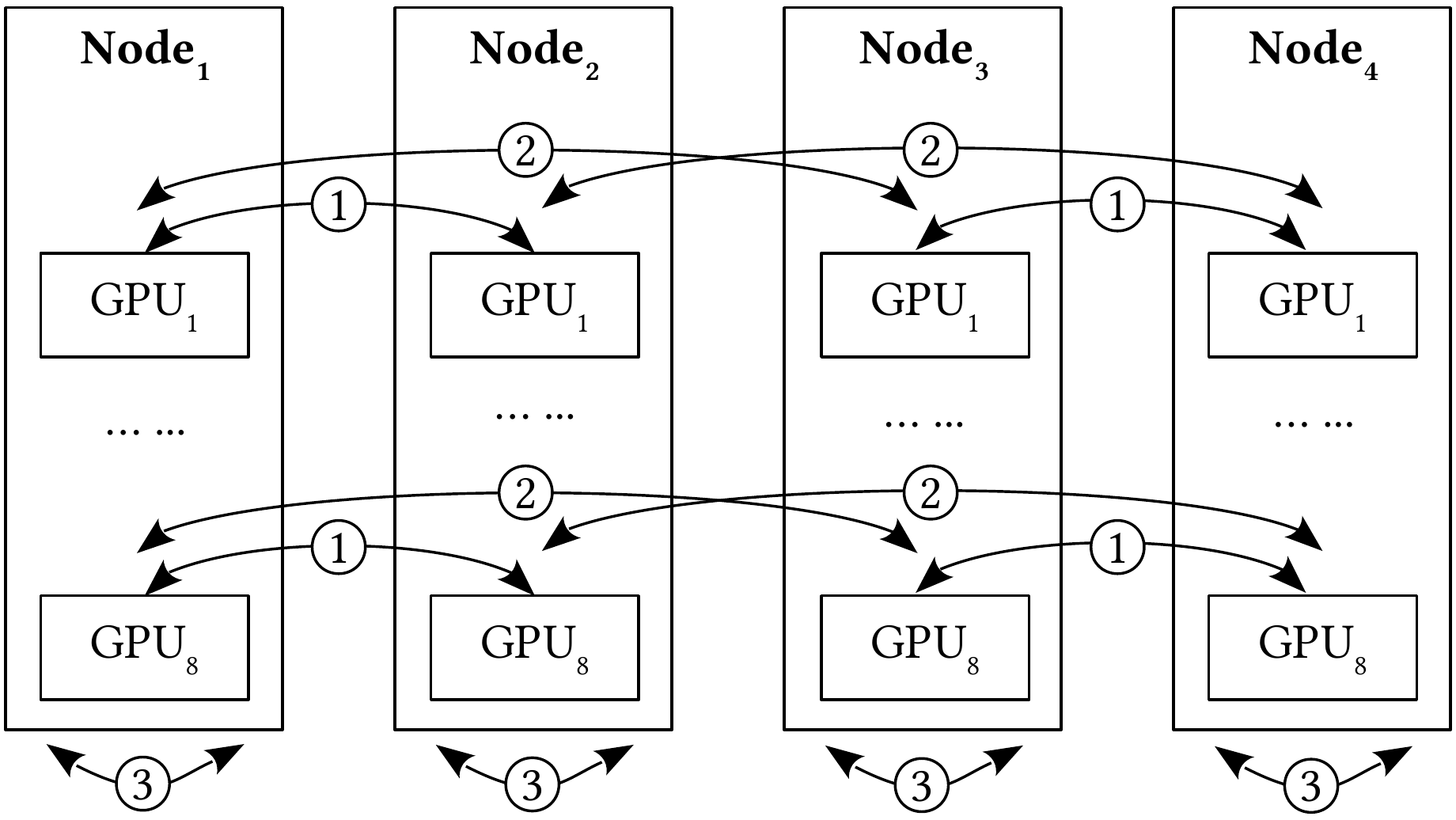}
  	\caption{All-reduce communication.}
  	\label{fig:collective}
\end{figure}

\subsection{Inter-Node GPU Communication}\label{sec:appendix:inter-comm}

\textbf{All-reduce communication.}
The parameter synchronization requires an all-reduce communication---each GPU needs to receive all parameter updates from other GPUs and then performs a reduction to accumulate these updates.
Figure~\ref{fig:collective} presents an example communication workflow. $4$ nodes are shown in this example. Each node contains $8$ GPUs.
Initially, the GPUs on $\textit{Node}_{1}$ exchange their parameter updates with their corresponding GPUs on $\textit{Node}_{2}$ (step {\small \circled{1}})---i.e., the $i^{\textit{th}}$ GPU on $\textit{Node}_{1}$ communicates with the $i^{\textit{th}}$ GPU on $\textit{Node}_{2}$.
Meanwhile, the GPUs with the same id on $\textit{Node}_{3}$ and $\textit{Node}_{4}$ share their data with each other.
Then, the GPUs on $\textit{Node}_{1}$ perform the communication with the ones on $\textit{Node}_{3}$ (step {\small \circled{2}}).
Likewise, the GPUs on $\textit{Node}_{2}$ and $\textit{Node}_{4}$ perform the same pattern communication in parallel.
After these two steps, each GPU on each node has collected all the parameter updates of its corresponding GPUs on other nodes.
An intra-node GPU tree all-reduce communication\footnote{\url{https://docs.nvidia.com/deeplearning/sdk/nccl-developer-guide/docs/usage/operations.html\#allreduce}} is executed to share the data across all $8$ GPUs on the same node (step {\small \circled{3}}).
Most of the communications are paralleled---$\log_{2}{\textit{\#nodes}}$ non-parallel inter-node and $\log_{2}{\textit{\#GPUs}}$ intra-node all-reduce communications are required to synchronize the parameters across all nodes.

\subsection{Dense Parameters}
As we discussed in the CTR prediction neural network example (Figure~\ref{fig:ctr-network}), besides the large-scale sparse parameters, there are a small number of dense parameters for the fully-connected layers. For any sparse input, all the dense parameters are referenced and updated. Therefore, we can pin these dense parameters in the HBM of all GPUs at the beginning of the training for better training performance. In extreme cases---we have insufficient HBM memory to replicate the dense parameters, we can shard the dense parameters as the sparse parameters and distribute them across all GPU HBMs.
The dense parameters are also synchronized as the sparse ones in the HBM-PS after each mini-batch is processed.

\section{MEM-PS Implementation}\label{sec:appendix:mem-ps}

\textbf{Cache policy.}
We target to cache the most recently and the most frequently used parameters in the memory to reduce SSD I/Os. In this paper, we leverage a cache eviction policy that combines two cache methods---Least Recently Used (LRU)~\cite{o1993lru} and Least Frequently Used (LFU)~\cite{sokolinsky2004lfu}.
Whenever we visit a parameter, we add it into an LRU cache. For the evicted parameters from the LRU cache, we insert them into an LFU cache. 
The evicted parameters from the LFU cache are collected---we have to flush them into SSDs before releasing their memory.
To guarantee the data integrity of our pipeline, we pin the working parameters of the current batch and the pre-fetched parameters of next iterations in the LRU cache---they cannot be evicted from the memory until their batch is completed.

\begin{figure*}[htbp]
	\includegraphics[width=.98\textwidth]{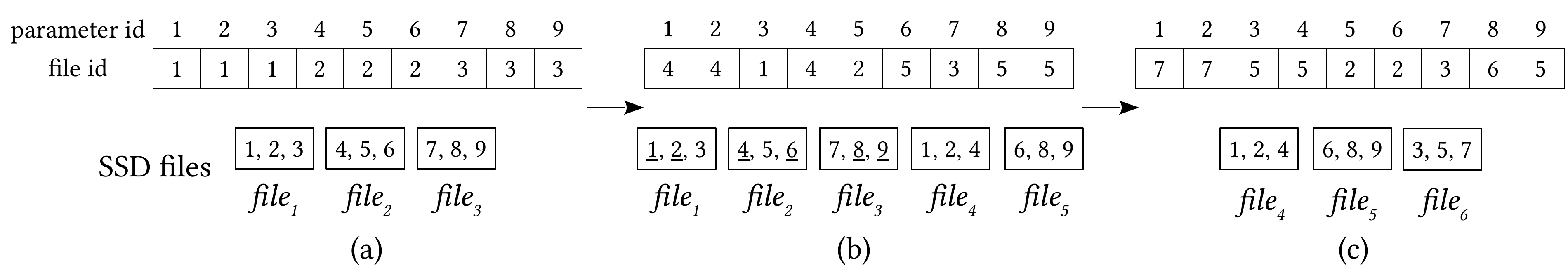}
  \vspace{-0.1in}
 	\caption{SSD-PS examples: (a) parameter-to-file mapping and parameter files; (b) $1,2,4,8,9$ are updated; (c) a compaction operation.}
  	\label{fig:ssd-example}
\end{figure*}

\section{SSD-PS Implementation}\label{sec:appendix:ssd-ps}
\textbf{Load parameters.}
The SSD-PS gathers requested parameter keys from the MEM-PS and looks up the parameter-to-file mapping to locate the parameter files to read. We have to read an entire parameter file when it contains requested parameters---a larger file causes more unnecessary parameter readings. This is a trade-off between the SSD I/O bandwidth and the unnecessary parameter reading---a small-size file cannot fully utilize the SSD I/O bandwidth. We tune the file size to obtain the optimal performance. Figure~\ref{fig:ssd-example}(a) depicts an example of parameter files on SSDs. In the example, each parameter file can store $3$ parameters.

\textbf{Dump parameters.}
Parameters evicted from the HBM-PS cache are required to be dumped onto SSDs. 
It is impractical to locate these parameters and perform in-place updates inside the original file---it poorly utilizes the SSD I/O bandwidth because it requires us to randomly write the disk.
Instead, our SSD-PS chunks these updated parameters into files and writes them as new files on SSDs---data are sequentially written onto the disk. After the files are written, we update the parameter-to-file mapping of these parameters.
The older versions of the parameters stored in the previous files become stale---these older values will not be used since the mapping is updated.
In Figure~\ref{fig:ssd-example}(b), we present an example for dumping parameters---$1,2,4,6,8,9$ are updated and dumped to SSD-PS. We chunked them into two files and write these two files onto the SSD---$\textit{file}_{4}$ and $\textit{file}_{5}$. The underlined values--the values of the updated parameters in the old files--are stale.

\textbf{File compaction.}
The SSD usage hikes as we keep creating new files on SSDs.
A file compaction operation is performed regularly to reduce the disk usage---many old files containing a large proportion of stale values can be merged into new files.
We adopt the leveled compaction algorithm of LevelDB~\footnote{\url{https://github.com/google/leveldb}} to create a lightweight file merging strategy. 
A worker thread runs in the background to check the disk usage. When the usage reaches a pre-set threshold, the SSD-PS scans the old parameter files, collects the non-stale parameters, merges them into new files, and erases the old files. The parameter-to-file mapping of the merged parameters is also updated in the file compaction operation.
Figure~\ref{fig:ssd-example}(c) illustrates the file compaction effects.
Before the compaction (Figure~\ref{fig:ssd-example}(b)), the stale values in $\textit{file}_{1}$, $\textit{file}_{2}$ and $\textit{file}_{3}$ occupy more than a half of the file capacity. We scan these files, merge the non-stale values into a new file ($\textit{file}_6$), and erase these files ($\textit{file}_{1}$ and $\textit{file}_{2}$).
The compaction operation may merge a large number of files on SSDs. In order to reduce the excessive merging, we set a threshold to limit the number of merged files---we only merge files that contain more than 50\% stale parameters. By employing this threshold, we can limit the total SSD space usage---the size of all parameter files will not exceed $2$ times ($1/50\%$) of the original non-stale parameter size.
Note that we do not need to read the entire file to obtain the proportion of stale parameters---a counter that counts the number of stale parameters is maintained as an auxiliary attribute for each file. When we update the parameter-to-file mapping, we accumulate the counter of the old file it previously maps to.

\section{Additional Related Work}\label{sec:appendix:relwork}
\textbf{In-memory cache management.}
Many caching policies have been developed for storage systems, such as the LRU-K~\cite{o1993lru}, DBMIN~\cite{chou1986evaluation}, LRFU\cite{lee2001lrfu}, and Semantic Caching~\cite{dar1996semantic}. These algorithms evict cache according to a combined weight of recently used time-stamp and frequency. In the web context, there is extensive work developed for variable-size objects. Some of the most well-known algorithms in this space are Lowest-Latency-First\cite{wooster1997proxy}, LRU-Threshold~\cite{abrams1995caching}, and Greedy-Dual-Size\cite{cao1997cost}. Unlike our caching problem, the parameter we tackle with has a fixed size and a clear access pattern in our CTR prediction model training---some parameters are frequently referenced. It is effective to keep those ``hot parameters'' in the cache by applying an LFU eviction policy. While our additional LRU linked list maintains the parameters referenced in the current pass to accelerate the hash table probing.

\textbf{Key-value store for SSDs.}
There is a significant amount of work on key-value stores for SSD devices. The major designs~\cite{andersen2009fawn,lim2011silt} follow the paradigm that maintains an in-memory hash table and constructs an append-only LSM-tree-like data structure on the SSD for updates. FlashStore~\cite{debnath2010flashstore} optimize the hash function for the in-memory index to compact key memory footprints. SkimpyStash~\cite{debnath2011skimpystash} moves the key-value pointers in the hash table onto the SSD. BufferHash~\cite{anand2010cheap} builds multiple hash tables with Bloom filters for hash table selection. WiscKey~\cite{lu2017wisckey} separates keys and values to minimize read/write amplifications. Our SSD-PS design follows the mainstream paradigm, while it is specialized for our training problem. We do not need to confront the challenges to store general keys and values. The keys are the index of parameters that distributes uniformly. It is unnecessary to employ any sophisticated hashing functions. Also, the values have a known fixed length, the serialized bucket on SSD exactly fits in an SSD block---I/O amplification is minimized. 

\end{document}